\documentclass[10pt]{article}

\usepackage{amsmath}
\usepackage{amssymb}
\usepackage{bm}
\usepackage{natbib}
\usepackage{graphicx}
\usepackage{enumerate}
\usepackage{hyperref}
\usepackage{url} 
\usepackage{caption}
\usepackage{subcaption}
\usepackage{tikz}
\usetikzlibrary{calc}
\usepackage{physics}
\onecolumn 

\begin{document}

\title{Relational event models with global covariates}

\author{Melania Lembo$^{a \footnote{Corresponding author: melania.lembo@usi.ch.}}$, R{\=u}ta Juozaitien{\.e}$^b \footnote{ruta.uzupyte@vdu.lt}$, Veronica Vinciotti$^c \footnote{veronica.vinciotti@unitn.it}$, Ernst C. Wit$^a \footnote{ernst.jan.camiel.wit@usi.ch}$ \\ \\
\small{$a.$ Institute of Computing, Universit\`a della Svizzera italiana, Lugano, Switzerland}\\
\small{$b.$ Faculty of Informatics, Vytautas Magnus University,Kaunas, Lithuania}\\
\small{$c.$ Department of Mathematics, University of Trento, Trento, Italy} \\
}
\maketitle

\abstract{
    Bike sharing is an increasingly popular mobility choice as it is a sustainable, healthy and economically viable transportation mode. By interpreting rides between bike stations over time as temporal events connecting two bike stations, relational event models can provide important insights into this phenomenon. The focus of relational event models, as a typical event history model, is normally on dyadic or node-specific covariates, as global covariates are considered nuisance parameters in a partial likelihood approach. As full likelihood approaches are infeasible given the sheer size of the relational process, we propose an innovative sampling approach of temporally shifted non-events to recover important global drivers of the relational process.\\
The method combines nested case-control sampling on a time-shifted version of the event process. This leads to a partial likelihood of the relational event process that is identical to that of a degenerate logistic additive model, enabling efficient estimation of both global and non-global covariate effects. The computational effectiveness of the method is demonstrated through a simulation study. The analysis of around 350,000 bike rides in the Washington D.C. area reveals significant influences of weather and time of day on bike sharing dynamics, besides a number of traditional node-specific and dyadic covariates.}
\maketitle

\section{Introduction}


    Analyzing bike share usage offers valuable insights into evolving mobility trends and the factors driving the adoption of this transportation mode. 
    The benefits of bike sharing include flexible mobility, emission reductions, increased physical activity, reduced congestion and fuel consumption, 
    individual financial savings, and support for multimodal transport connections \citep{Shaheen_etal_2010_bikeshare}. 
    Studies have shown that bike sharing is increasingly utilized for both commuting and recreational purposes, 
    reflecting its versatility and growing appeal as a practical urban mobility option \citep{Fishman_etal_2013_bikeshare_rev_old}. 
    In cities such as New York and Washington, D.C., bike share users are more likely to engage in short trips, particularly for first-mile/last-mile connections to public transit.
    The rise in bike share membership and the increased frequency of trips highlight the impact of infrastructure investments, such as the expansion of bike lanes and station networks, which help create a more supportive environment for cycling in urban areas.

Weather conditions significantly influence bike sharing behavior, with extremely low or high temperatures and heavy precipitation being associated with a decrease in usage \citep{Gebhart_Noland_weather_bikes}. Time of day is another crucial factor, as daylight and working hours play a role in the decision to ride a bike \citep{hampshire2012analysis_tod}.
    Understanding these trends is essential for urban planners, policymakers, and bike sharing providers who aim to develop efficient and sustainable transportation networks. 
    The temporal nature of this phenomenon is critical, and when viewed as a sequence of interactions, i.e., bike rides, between a set of entities, i.e., bike stations, over time, bike sharing can be conceptualized as a dynamic network.
    We will focus on publicly available data from Capital Bikeshare, consisting of 350,000 rides in Washington D.C. during July 2023.

Relational event models (REMs) provide an effective framework for modelling general dynamic networks \citep{remreview}. As such, they have been used for modelling networks evolving over time in a number of applications where time-stamped interactions are available, such as radio communications \citep{Butts2008}, email communications \citep{perrywolfe2013}, interactions of users with online learning platforms \citep{Vu2015},
invasions of countries by alien species \citep{juozaitiene2023analysing}, patent citations \citep{filippimazzola2023stochastic} and financial transactions \citep{Bianchifinancial}.

Relational event models represent interactions as a marked counting process, where the mark is the edge indicator and the count the number of times that the edge has occurred up to that point in time. Relational event models have been extended to the case of signed, weighted \citep{Brandes2009NetworksES}  and polyadic, i.e., one-to-many, relations \citep{perrywolfe2013,RHEM_Lerner_2023}. Covariates can be either exogenous, i.e., representing intrinsic attributes or characteristics of the entity or interaction, or endogenous, i.e., with values that depend on the evolution of the network. The effects of potential drivers, originally included in the model as fixed effects, have been extended to random \citep{uzaheta2023random,boschi2023smooth}, non-linear and time-varying effects \citep{Juozaitiene_nodal_2024,boschi2023smooth}. 

All of the effects considered above are essentially node-specific or dyadic. The main reason for this is that the predominant REM inference paradigm employs the partial likelihood, in which any global effect, such as the baseline hazard, is treated as a nuisance parameter that cancels out in the partial likelihood. In order to infer global effects, one option is to consider the full likelihood. However, the exact full likelihood is intractable, as it involves an integral over the linear predictor between the event times.  \citep{dynam_original_paper} and \cite{stadtfeld_Block_global_FL} propose an approximation, by assuming that the linear predictor is constant between event times, thus reducing the integral to a simple multiplication. However, even with this simplification, the likelihood calculation scales quadratically with the number of nodes, which is prohibitive for anything but very small networks --- and certainly would not work for the bike sharing data, containing over a million of bike station pairs. Alternatively, in a partial likelihood framework, the overall global effect, also called the cumulative baseline hazard, can be estimated a posteriori \citep{breslow1972, kalbfleischprentice, rutasplines}. However, this does not provide a way to deal with general global covariates. 
\cite{kreissglobal} suggest using a two-step profile likelihood approach, but this tends to be computationally challenging.
Finally, there are examples in the literature that ostentatiously seem to be inferring global effects with the partial likelihood. For example, \cite{amati_2019} considers a temporal day-of-week effect,  but this is included in the model only as interaction effects with dyadic statistics, such as reciprocity. Interaction effects of global effects with dyadic effects are effectively dyadic and are identifiable with the ordinary partial likelihood. So these methods cannot be used to estimate truly global effects.

In the case of the bike sharing study, besides considering node-specific and dyadic covariates, e.g., local density of bike-stations and distance between bike stations, an important consideration are global covariates, i.e., covariates that are time-dependent but constant for all interacting pairs. Global covariates, such as weather and time of day, are particularly relevant when modelling the rate of bike share. In this paper, we propose a method that is based on the partial likelihood and that is able to estimate both global and non-global effects. The method is exact, as it does not require the piecewise constant assumption, such as in the full likelihood approaches mentioned above. The idea that we develop is that of considering a time-shifted version of the original event process. This leads to a partial likelihood in which the contribution from global covariates does not cancel out, as the observations in the risk set will have received different shifts and will therefore be evaluated at different time points. Standard inference using this time-shifted partial likelihood such as those currently used for REMs will naturally apply also in this case. For small to medium dynamic networks the proposed method is effective. As the computation of the partial likelihood scales quadratically with the number of nodes, the method can be combined with nested case-control sampling \citep{borgan95, Vu2015} in order to make it feasible for essentially infinitely large networks. Unlike the approximate full likelihood approach, the method is consistent. Moreover, if only one non-event is sampled for each event, the partial likelihood can be reformulated as that of a degenerate logistic regression model, allowing the inclusion of non-linear smooth additive terms and the use of existing efficient techniques for statistical inference \citep{woodgeneralized, boschi2023smooth,filippimazzola2023stochastic}. The result is a computationally efficient inferential approach for REMs involving a potentially large number of entities, whose rate of interaction depends smoothly, but not necessarily linearly, on node-specific, dyadic or global covariates.

In section \ref{sec:bike_data_descr}, we describe the bike sharing data for Washington D.C.. Section \ref{sec:methods} gives a description of REMs using multivariate counting processes and extends the model formulation with the inclusion of global covariates. Section \ref{sec:inference} presents the proposed time-shifted event process for statistical inference under a partial likelihood framework. It derives the degenerate logistic additive logistic regression model that results from the use of nested case-control sampling. Section \ref{sec:simulation} evaluates the proposed methodology via a simulation study. In section \ref{sec:bike_data} we return to the bike sharing study and show the identified drivers of the bike sharing process in Washington D.C. in July 2023. 

\section{Bike sharing as a relational process}\label{sec:bike_data_descr}

Bike sharing has increased considerably over the past decade. As of 2019, over 2000 cities worldwide provide a public bike sharing service \citep{Fishman_bikeshare_book}. The popularity of this mobility option, on one hand, contributes to improved population health, as a result of the increase in physical activity, and, on the other hand, addresses climate change concerns through air and noise pollution reduction. Another advantage of this system is in its offer of an easy one-way travel option: it grants users the flexibility to choose a different mobility option on the way back and it represents an effective way to access alternative public transportation options (such as buses or trains) for trips requiring multi-modal transportation \citep{Fishman_bikeshare_rev}. A survey from 2016 conducted by LDA consulting for Capital Bikeshare, a bike sharing provider in Washington D.C., reports that 57\% out of the 6,395 respondents were motivated to purchase a membership in order to have access to one-way travel options. Furthermore, that same report states that 71\% of respondents used bike sharing at least occasionally to access alternative public transportation options \citep{CBS_2016_report}.

GPS technologies, which are widely adopted by public bike sharing services, provide an easy access to time-stamped information on bike rides occurring between certain locations. This can be used to investigate what aspects may explain bike-sharing dynamics. Bike rides between stations can be conceptualized as a relational process. Each trip represents a directed link between two nodes --- stations --- indicating a flow of users through the city.  By viewing these rides as a process that evolves in time, dynamic network techniques, presented in the following sections, can be used to analyze how aspects such as weather, geographical distance or time of day, drive bike sharing behavior. This can provide important insights into system performance, rider behavior and urban dynamics, which can be used by urban planners and bike sharing providers for future infrastructure investments and revenue optimization.

In this paper, we focus on analyzing bike sharing data for the city of Washington D.C., which is available publicly from \url{https://capitalbikeshare.com/system-data}. In particular, we consider the period July 9th-31st, 2023, which involves around 350k bike rides among more than 1300 locations in the area. 
Weather information is taken from \url{https://www.wunderground.com/} and assumed the same throughout the entire service area. Information about distances between stations  is obtained from Open Street Maps using the \texttt{R} package \texttt{osrm}, using the known latitude/longitude coordinates of the locations. The locations are the nodes of the network, while an edge goes from a station $s$ to another station $r$ at time $t$ if a bike is taken from station $s$ at time $t$ and is then left at station $r$. Details of the pre-processing are available in the Supplementary Materials and the related code is available from \href{https://github.com/MelaniaLembo/Global-covariates-REM.git}{\texttt{https://github.com/MelaniaLembo/Global-covariates-REM.git}}.

The purpose of a specific bike ride can be related to a work commute or to a personal travel need \citep{Fishman_etal_2013_bikeshare_rev_old,CBS_2016_report}. While information on ride-specific purposes is not available, temporal patterns, such as how usage varies by time of day, e.g., whether it increases during working hours and daylight, can be measured directly. Gaining insights on this can be critical for managing bike fleet availability. It is reported extensively in transportation-related literature that convenience to one's need is an important factor in the choice to use a bike sharing service \citep{Fishman_bikeshare_book}. One feature of this relates to station placement: easy and close access to a bike station promotes usage of the service. More in general, aspects related to stations density per area should be investigated in order to understand how they affect the choice of riding a bike between two stations. Finally, weather conditions clearly plays an important role, as temperature and precipitation can have an effect on the frequency of rides.

\begin{figure}[tb]
    \begin{subfigure}{.49\linewidth}
         \centering
            \includegraphics[scale=0.25]{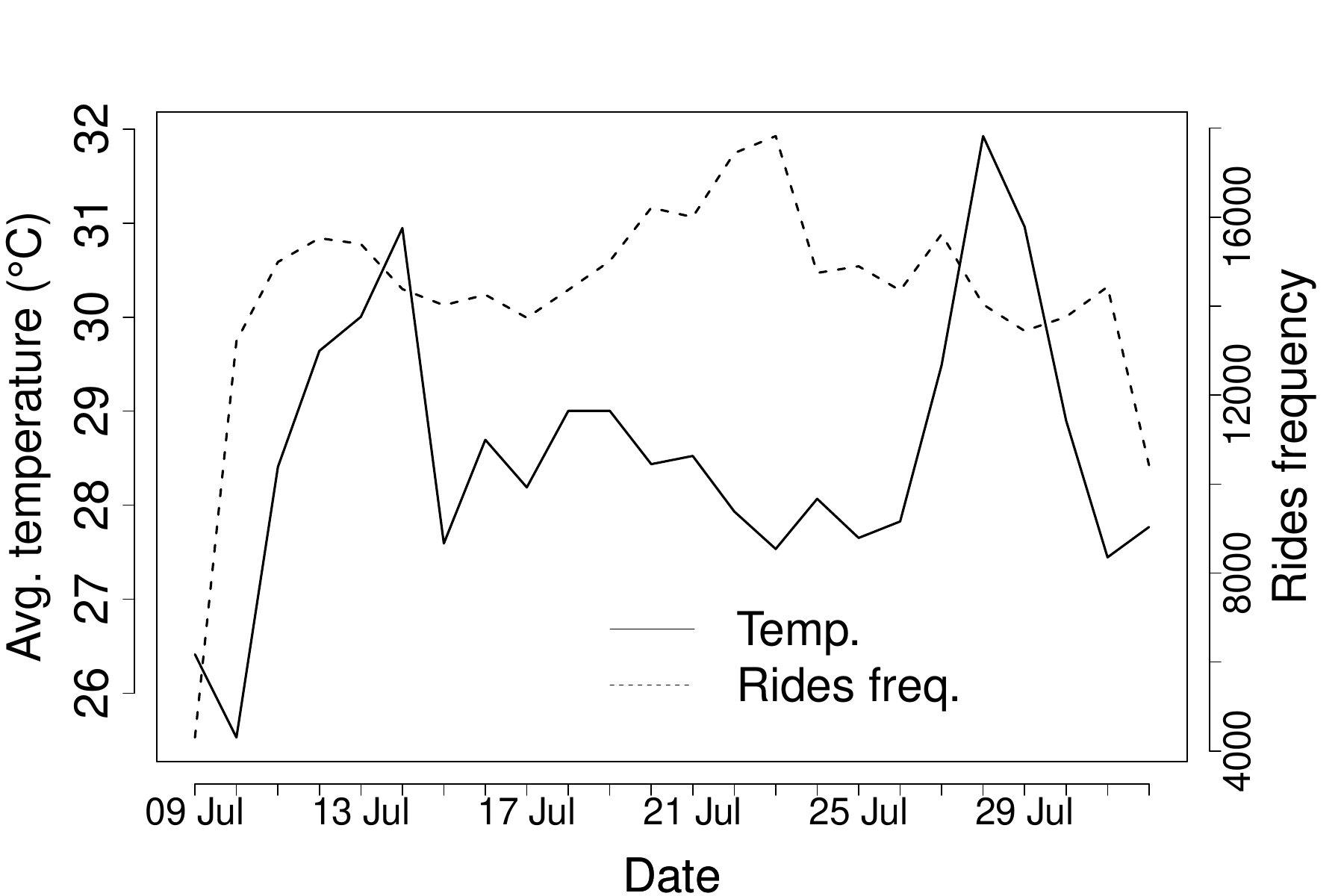}
        \caption{}
        \label{fig:EDA_bike_temp}
    \end{subfigure}
    \hfill
    \begin{subfigure}{0.49\linewidth}
        \centering				
         \includegraphics[scale=0.25]{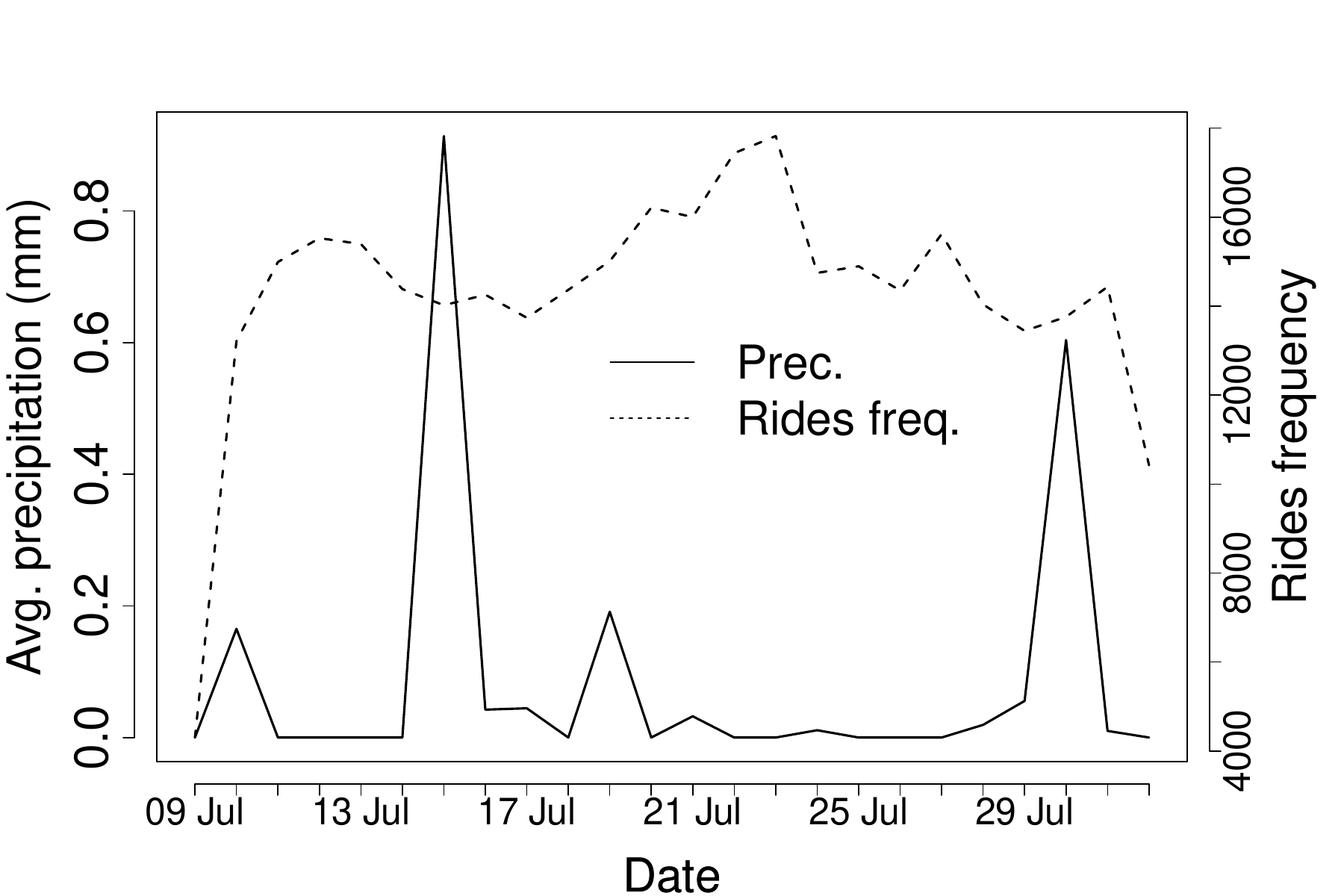}
         \caption{}
         \label{fig:EDA_bike_prec}
    \end{subfigure}
    \caption{(a) Average daily temperature and (b) average daily precipitation relative to the daily number of rides (dashed line, y-axis on the left) in Washington D.C. during July 9th-31st, 2023. A decrease in rides frequency generally aligns to days of lower temperatures and heavier precipitations (July 14-18th and toward the end of the month).}
\label{fig:EDA_bike_temp_prec_freq}
\end{figure}

\begin{figure}[tb]
    \begin{subfigure}{.49\linewidth}
         \centering
                    \includegraphics[scale=0.25]{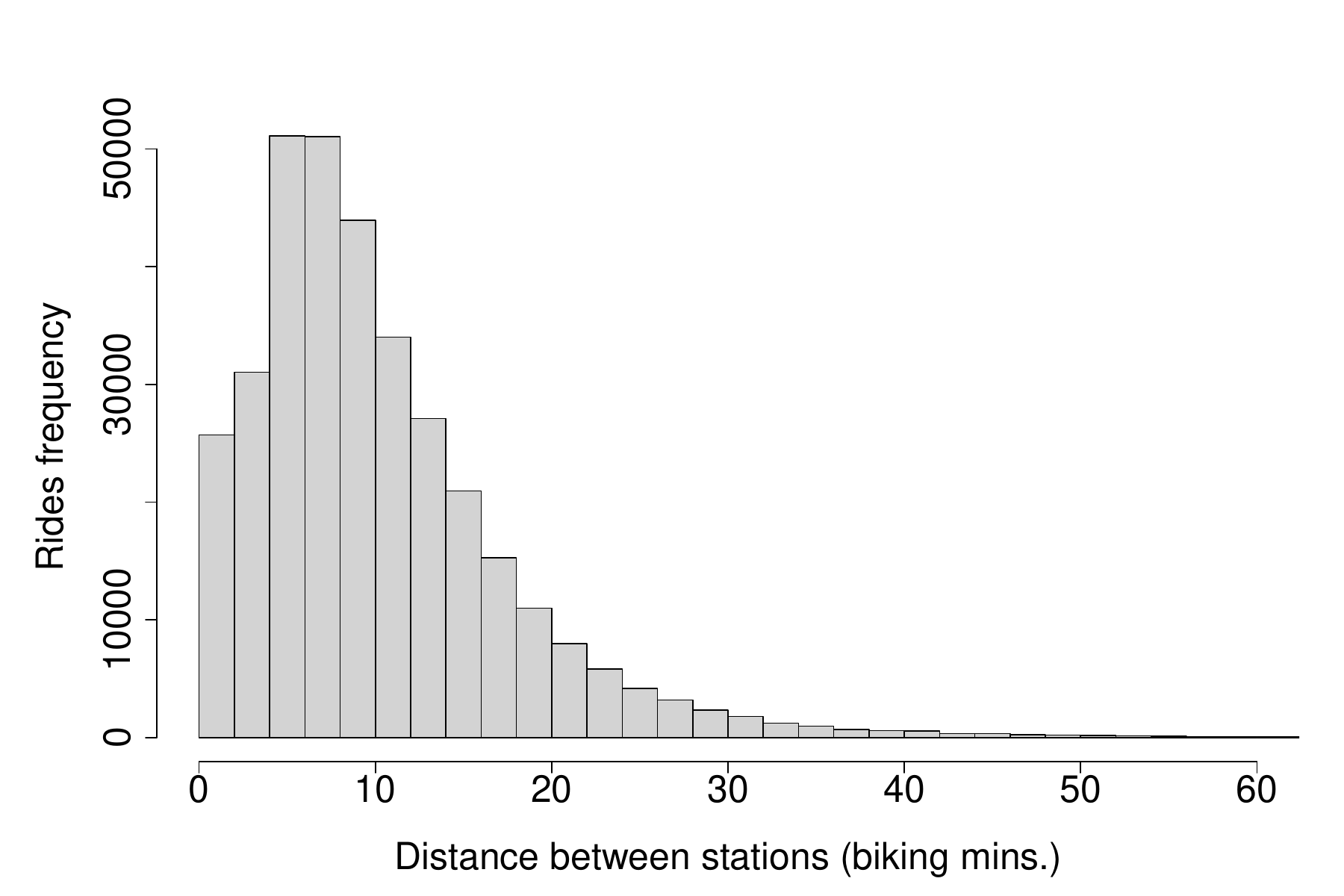}
        \caption{}
        \label{fig:EDA_bike_dist}
    \end{subfigure}
    \hfill
    \begin{subfigure}{0.49\linewidth}
        \centering				
         \includegraphics[scale=0.25]{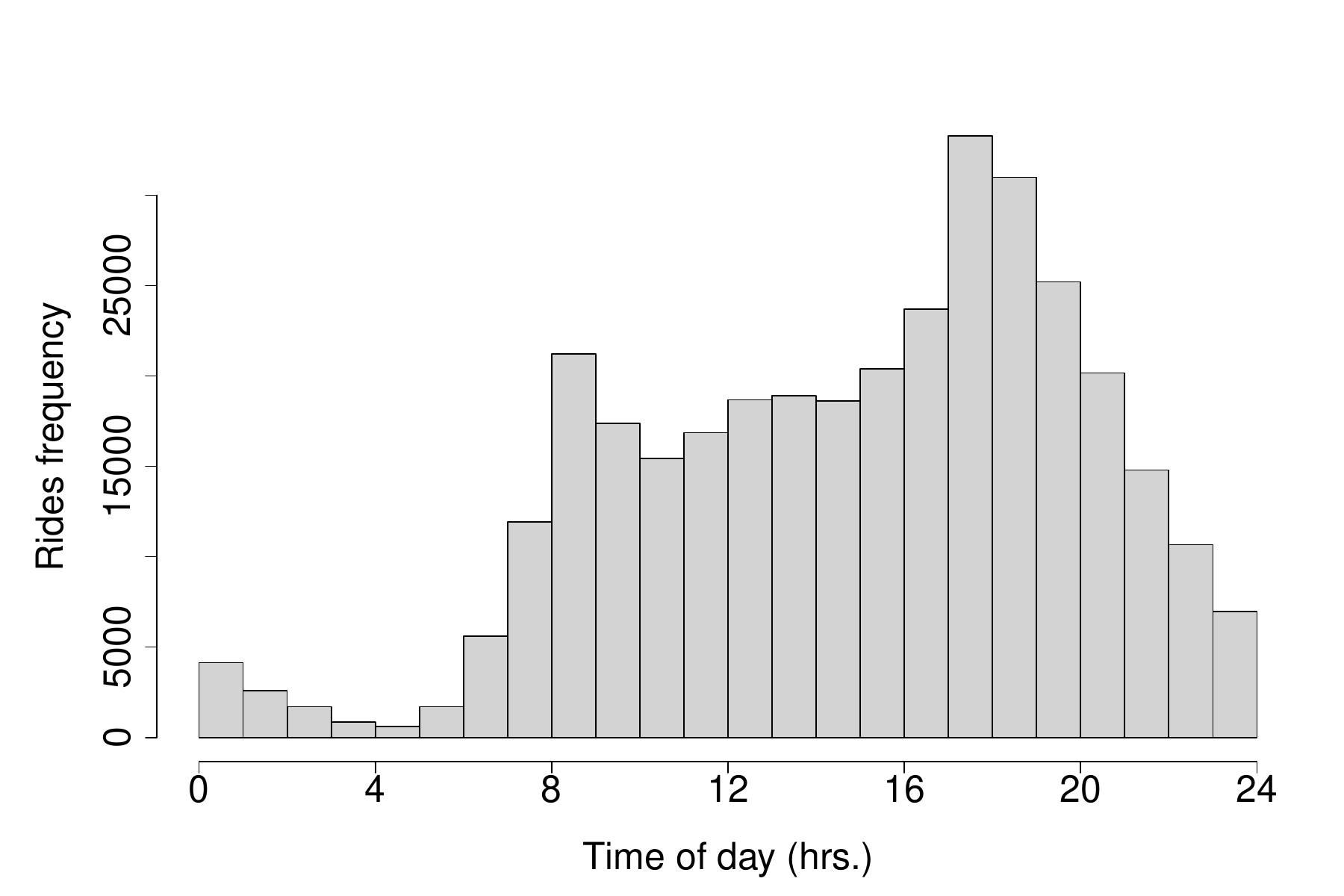}
         \caption{}
         \label{fig:EDA_bike_tod}
    \end{subfigure}
    \caption{Distance of rides (a) and time of day (b) in relation to ride frequency: (a) Short trips, up to 20 minutes biking distance, are the most frequent. (b). Higher frequency of rides is observed during the day and in particular related to beginning (4-9 am) and end (around 6 pm) of working hours.}
\label{fig:EDA_bike_dist_tod}
\end{figure}

Figures~\ref{fig:EDA_bike_temp} and \ref{fig:EDA_bike_prec} show the average daily temperature and average daily precipitation, respectively, relative to the number of rides during the chosen period of observation. One can notice a general decrease on the number of rides between July 14th-18th and at the end of the month, coinciding with a decrease in temperature and an increase in the precipitation level during these period. Furthermore, Figure~\ref{fig:EDA_bike_dist} shows that bike shares are mostly used for short trips, with the most frequent ones being around 10 minutes long, and that the number of rides decreases considerably as the distance between bike stations -- measured in biking minutes -- increases: only 0.05\% ($\approx$ 170) of the observed rides correspond to a distance that is greater than one hour. Finally, Figure~\ref{fig:EDA_bike_tod} shows that daylight and working hours have an impact on bike share usage with fewer rides observed at night and with a peak of usage at 9am and around 6pm.

\section{Relational event model with global covariates}
\label{sec:methods}

A relational event is a directed interaction $s \rightarrow r$ between a sender $s$ and a receiver $r$,
occurring at a specific point in time. Formally, it can be described by a triplet
\[
    (t,s,r) \in [0,T] \times \mathcal{S} \times \mathcal{C},
\] 
where $\mathcal{S}$ is the set of senders, 
$\mathcal{C}$ the set of receivers and $[0,T]$ a bounded time interval. We consider a sequence of $n$ relational events happening in $[0,T]$, as a realization from a market point process
\begin{equation}
    \label{eq:mpp}
    \mathbf{M} = \{\left(t_k,(s_k,r_k)\right) \ | \ k \ge 1\}.
\end{equation}
with mark space $\mathcal{S} \times \mathcal{C}$, where $\{t_k\}_{k=1}^n$ are the ordered and distinct event times at which the corresponding interaction $s_k \rightarrow r_k$ occurs. Associated with this marked point process, there is a multivariate counting process $\mathbf{N} = \{N_{sr}\}_{(s,r) \in \mathcal{S} \times \mathcal{C}}$ that records the number of occurrences of all interactions $(s,r)$
in $[0,t]$,
\begin{equation}
    \label{eq:mcp}
    N_{sr}(t)= \sum_{k\ge 1}\mathbb{I}(t_k \le t,\ s_k = s,\ r_k = r). 
\end{equation}
The assumption that the time points $t_k$ are distinct guarantees that no two edges occur at the same time. Furthermore, we assume that each component $N_{sr}$ has finite expectation, satisfies the Markov property and that no events occur at time 0. Since any adapted, non-negative and non-decreasing process with finite expectation is a sub-martingale, by the Doob-Meyer decomposition theorem, there exist an $\mathcal{H}_t$-predictable process $\Lambda_{sr}$ such that
\[ N_{sr}(t) = M_{sr}(t) + \Lambda_{sr}(t), \]
where $\mathcal{H}_t$ is the history of the process up to time $t$, $M_{sr}(t)$ is an $\mathcal{H}_t$-martingale and the predictable part $\Lambda_{sr}(t)$ is given by
\[
\Lambda_{sr}(t) = \int_0^t \lambda_{sr}(u)du,
\]
under the condition that the waiting times distribution are absolutely continuous. The intensity process $\lambda_{sr}(t)$ represents the rate of event $(s,r)$ occurring at time $t$.

REMs describe the dependence of the intensity process on  covariates of interest via a semi-parametric Cox model \citep{coxph1972}. Whereas existing approaches specify the parametric part of the model via node-specific or dyadic variables, the aim of this paper is to include global covariates, i.e., covariates that describe system-wide characteristics that are time-dependent but constant across all interacting pairs. We extend the existing formulation of the REM intensity process as follows, 
\begin{equation}
    \label{eq:lambda_sr_gc}
    \lambda_{sr}(t) = Y_{sr}(t)\lambda_0\text{exp}\left\{\sum_{l=1}^{q} f_l\biggl(\mathbf{x}^{(l)}_{sr}(t)\biggr) + g_0(t) + \sum_{h=1}^{w} g_h\biggl(\mathbf{x}^{(h)}(t)\biggr)\right\}.
\end{equation}
Here $Y_{sr}(t)$ is an $\mathcal{H}_t$-measurable indicator taking value 1 if the pair $(s,r)$ is at risk of occurring at time $t$, and 0 otherwise, and the effect of covariates is split between the $q$ non-global covariates $\mathbf{x}^{(l)}_{sr}$ (e.g. distance between stations), which enter the model via $q$ arbitrary, possibly non-linear, smooth functions $f_l$, and $w$ global covariates $\mathbf{x}^{(h)}$ (e.g. weather-related variables) with their associated functions $g_h$. Any residual time-dependence not accounted for by the global and non-global covariates effects, is modelled through an arbitrary non-negative global time effect $g_0$. Combined with the constant $\lambda_0$, the function $\lambda_0 e^{g_0(t)}$ plays a role similar to that of the baseline hazard, used in event history modelling. However, there is a difference, as the traditional baseline hazard aggregates  all  global effect terms $g$ into one time-dependent function. To distinguish the two, we later refer to this component of our model as the global time effect. The processes associated to the risk set $\mathcal{R}(t) = \{(s,r)\in \mathcal{S} \times \mathcal{C} \ | \ Y_{sr}(t) = 1\}$ and all the covariates are assumed to be left-continuous and adapted to $\mathcal{H}_t$.

\section{Inference}
\label{sec:inference}

In this section we present an extension of existing inferential partial likelihood techniques, that allow for the estimation of the global effects $\{g_h\}$ and the global time effect $g_0$ in (\ref{eq:lambda_sr_gc}). We  construct  a time-shifted version of the original event process (\ref{eq:mpp}). Although the partial likelihood in this time shifted-process allows for consistent estimation of the global effects, it can be computationally demanding. For this reason, we apply nested case-control sampling on this newly defined process and construct the resulting partial likelihood. We conclude this section by showing that, in the case of only one non-event sampled for each event, the resulting partial likelihood is that of a degenerate logistic model. This opens up the possibility for flexible and efficient inferential techniques used in additive logistic modelling.

\subsection{Time-shifted event process}
\label{subsec:shifted_process_def}
We construct a new event process $M^e$ from the original marked point process (\ref{eq:mpp}) by shifting the marks $(s,r)\in \mathcal{S}\times \mathcal{C}$ by independent positive shifts. More precisely, we consider a collection of random shift variables $\mathbf{H}=\{H_{sr}\}_{(s,r) \in \mathcal{S}\times\mathcal{C}}$, independent of $M$ such that the random variables $H_{sr}$ have non-negative support. Subsequently, we define the shifted marked point process associated with $\mathbf{M}$ as 
\[ M^e = \{ \left(t_j+H_{s_jr_j}, (s_j, r_j)\right) ~|~ (t_j,s_j,r_j) \in M \}.\]
Related to this marked point process, there exist a multivariate counting process $\mathbf{N^e} = \{N_{sr}^e\}_{(s,r) \in \mathcal{S}\times\mathcal{C}}$, with the following properties,
\begin{align}
\label{eq:Neprocess}
    N_{sr}^e(t) = \begin{cases}
        0, & t \in [0,H_{sr}) \\
        N_{sr}(t-H_{sr}), & t \in [H_{sr},H_{sr} + T] \\
        N_{sr}(T), & t \in (H_{sr} + T, T^e],
    \end{cases}
\end{align}
where $T^e = T + \max_{(s,r) \in \mathcal{S}\times\mathcal{C}}H_{sr}$. This represents a submartingale, which allows a new Doob-Meyer decomposition, whereby the cumulative hazard $\Lambda_{sr}^e(t) =\int_0^t \lambda_{sr}^e(u)~du$ is adapted to a new filtration $\mathcal{F}_t$. Intuitively, at time $t$, $\mathcal{F}_t$ contains information on $\mathbf{H}$ and the evolution of $\mathbf{N}$, through the $\sigma$-algebras $\mathcal{H}_{t-H_{sr}}$ for all the different pairs $(s,r)$. Conditional on the shift process, say $H_{sr} = h_{sr}$ for all $(s,r) \in \mathcal{S}\times\mathcal{C}$, the hazard of a pair $(s,r)$ occurring at time $t$ for this shifted process is the same as the hazard of the original event process evaluated at time $t-h_{sr}$, 
\begin{equation}
	\label{eq:shifted_lambda_sr}
	\lambda_{sr}^e(t)  = \mathbb{I}(h_{sr}\leq t \leq h_{sr} + T) \lambda_{sr}(t - h_{sr}).
\end{equation}

Using (\ref{eq:shifted_lambda_sr}) and the independence assumption between $\mathbf{H}$ and $M$, we have that the probability of the interaction 
$(s,r)$ occurring at a shifted time $t$, given $\mathcal{F}_{t-}$ and that an event happened at that time, follows a multinomial distribution over the elements of $\mathcal{R}^e(t)$, with probabilities that only depend on the intensity process of the original event process $\mathbf{N}$, namely
\begin{align}
    P\left((s,r) \mbox{ happens in } M^e \mbox{ at }t ~|~\mbox{event at } t; \mathbf{f},\mathbf{g}\right) =  \frac{\lambda_{sr}(t-h_{sr} )}{\displaystyle \sum_{(s^*,r^*) \in \mathcal{R}^e(t)} \lambda_{s^*r^*}(t-h_{s^*r^*})},
\end{align}
where the dependence on $\mathbf{f} =(f_1,\dots,f_q)$ and $\mathbf{g} =(g_0,g_1,\dots,g_w)$ is through the definition of $\lambda_{sr}$ in (\ref{eq:lambda_sr_gc}).
If we then consider a realization of $n$ events of the process $M^e$, 
$\{(t_1^e,s_1^e,r_1^e), \dots, (t_n^e, s_n^e, r_n^e)\},$ 
under the conditional independence assumption, the partial likelihood  becomes
\begin{align}
    \label{eq:full_pl}
    \mathcal{L}^P(\lambda_0, \mathbf{f},\mathbf{g}) = \prod_{j=1}^{n} \frac{\lambda_{s_j^er_j^e}(t_j^e-h_{s_j^er_j^e})}{\displaystyle \sum_{(s^*,r^*) \in \mathcal{R}^e(t_j^e)} \lambda_{s^*r^*}(t_j^e-h_{s^*r^*})},
\end{align}
where the risk set $\mathcal{R}^e(t) = \{(s,r)\in \mathcal{S} \times \mathcal{C} \ | \ Y_{sr}(t-h_{sr}) = 1\}$.
Crucially, this partial likelihood is informative in terms of the effects of all covariates, including those that are only time dependent, as the intensities of each event and all the other interactions in the risk set are now all a.s. evaluated at different time points.

\subsection{Nested case-control sampling on shifted event process}
\label{subsec:modified_NCC}

Similarly to the traditional partial likelihood,  our proposed time-shifted version in (\ref{eq:full_pl}) presents computational challenges already for networks of a moderate size. Indeed, the denominator involves the sum over all pairs at risk of interacting at a particular point in time and therefore scales quadratically with the number of nodes.  We address this challenge with the use of nested case-control sampling, whereby a number of non-events are uniformly sampled from the risk set at a specific event time \citep{borgan95}.  

We will focus on the case where, at a general shifted event time $t$ corresponding to an event $(s,r)$, 
a single non-event $(s^*,r^*) \neq (s,r)$  is sampled at random from the risk set $\mathcal{R}^e(t)$ at that time. The marked point process is then extended to
\begin{align}
    \label{eq:ex_mpp}
    \{(t_j^e, (s_j^e, r_j^e,\tilde{\mathcal{R}}_{j}^{e} )) \ | \ j \ge 1\},
\end{align}
where $\tilde{\mathcal{R}}_{j}^{e}$ denotes the sampled risk set containing the non-event sampled from $\mathcal{R}^e(T_j^e)$ and the event $(s_j^e, r_j^e)$ occurred at time $T_j^e$. This process is no longer adapted to $\{\mathcal{F}_t\}_{t\geq 0}$ due to the additional variability originating from the risk set sampling. We therefore work with an augmented filtration generated by both the process and the sampling, namely $\{\mathcal{G}_t\}_{t\geq 0}$ with $\mathcal{G}_t = \mathcal{F}_t \vee \sigma\big(\tilde{\mathcal{R}}_{j}^{e} \ ; \ T_j^e \leq t \big)$. With the independent sampling assumption, the intensity processes $\boldsymbol{\lambda}^e$ of $\mathbf{N}^e$ in (\ref{eq:Neprocess}) remain the same also with respect to this augmented filtration. 



Associated with (\ref{eq:ex_mpp}), we consider the multivariate counting process of components
\[
    N_{sr,R}(t) = \sum_{t^e_j\leq t} \mathbb{I}\left(s_j^e=s,r_j^e=r, \tilde{\mathcal{R}}_{j}^{e} =R \right),
\]
which counts the observed number of interactions $(s,r)$ in $[0,t]$ with corresponding sampled risk set $R$ made of $(s,r)$ and the sampled non-event $(s^*,r^*)$.
The $\mathcal{G}_t$-intensity process is then obtained by multiplying that of $\mathbf{N}^e$ with the uniform sampling distribution for the non-event, that is
\begin{align}
\label{eq:lambda_NCC}
\lambda_{sr,R}(t) = \lambda_{sr}^e(t) \dfrac{\mathbb{I}(R \subset\mathcal{R}^{(e)}(t), (s,r) \in R, |R| = 2)}{n^{e}(t)-1},
\end{align}
with $n^e(t) = |\mathcal{R}^{(e)}(t)|$.
From this, it follows that the probability of $(s,r)$ occurring at time $t$, conditioned on $\mathcal{G}_{t-}$ and that either the event $(s,r)$ or non-event $(s^*,r^*)$ can happen at time $t$, is given by 
\begin{align}
    \label{eq:pi_NCC}
    P\left((s,r)|R; \mathbf{f},\mathbf{g}\right)    &= \frac{\lambda_{sr,R}(t)}{\lambda_{sr,R}(t) + \lambda_{s^*r^*,R}(t)}.
\end{align}

Consider the $n$ events from (\ref{eq:ex_mpp}) and denote these with $(t_j^e, (s_j^e, r_j^e,R_{j}^{e} )) $ for $ j = 1, \dots ,n $. Since
the elements in $R_{j}^{e}$ are equally likely, the uniform probability in (\ref{eq:lambda_NCC})  evaluated at each of these events
cancels out in (\ref{eq:pi_NCC}), leading to a sampled partial likelihood that depends only on the intensity processes of $\mathbf{N}$, namely
\begin{align*}
    \mathcal{L}^{PS}(\lambda_0,\mathbf{f},\mathbf{g}) = \prod_{j=1}^{n} \frac{\lambda_{s_j^er_j^e}(t_j^e - h_{s_j^{e}r_j^{e}})}{\lambda_{s_j^{e}r_j^{e}}(t_j^e - h_{s_j^{e}r_j^{e}}) + \lambda_{s_j^{e*}r_j^{e*}}(t_j^e - h_{s_j^{e*}r_j^{e*}})}.
\end{align*}
Note that depending on the shift distribution and the evolution of the risk set of the original process, it could happen that, at some shifted event times, the risk set of the shifted process is only composed of the actual event. In this case it is not possible to sample a non-event, rendering this observation uninformative in the sampled partial likelihood above. For all other observations, considering the $k \in \{1,\dots, n\}$ such that $t_k=t_j^e - h_{s_j^{e}r_j^{e}}$ and $(s_j^e,r_j^e) = (s_k,r_k)$, which exists by construction of the shifted process, and similarly $t_k^* = t_j^e - h_{s_j^{e*}r_j^{e*}}$ and $(s_k^*,r_k^*)$ for the non-event, we can rewrite the sampled partial likelihood more conveniently as 
\begin{align}
    \label{eq:final_PL}
    \mathcal{L}^{PS}(\lambda_0,\mathbf{f},\mathbf{g}) = \prod_{k=1}^{n} \frac{\lambda_{s_kr_k}(t_k)}{\lambda_{s_kr_k}(t_k) + \lambda_{s_k^{*}r_k^{*}}(t_k^*)}.
\end{align}

\subsection{Degenerate logistic additive modelling}
\label{subsec:add_model}
By substituting (\ref{eq:lambda_sr_gc}) in the sampled partial likelihood (\ref{eq:final_PL}) and dividing both numerator and denominator by $\lambda_{s_k^*r_k^*}(t_k^*)$ we obtain the likelihood of a degenerate logistic regression model,
\begin{align}
    \label{eq:bin_gam_PL}
    \mathcal{L}^{PS}(\mathbf{f},\mathbf{g}) = \prod_{k=1}^{n} \frac{\exp\big\{  \boldsymbol{\Delta_k}(\mathbf{f}; \mathbf{x}_{s_kr_k}) + \boldsymbol{\Delta_k}(\mathbf{g}; \mathbf{x}_k)\big\}}{ \exp\big\{ \boldsymbol{\Delta_k}(\mathbf{f}; \mathbf{x}_{s_kr_k}) + \boldsymbol{\Delta_k}(\mathbf{g}; \mathbf{x}_k)\big\}+ 1},
\end{align} 
where
\begin{align*}
    \boldsymbol{\Delta_k}(\mathbf{f}; \mathbf{x}_{s_kr_k}) &=  \sum_{l=1}^{q} \big[f_l\big(\mathbf{x}^{(l)}_{s_kr_k}(t_k)\big) - f_l\big(\mathbf{x}^{(l)}_{s_k^*r_k^*}(t_k^*)\big)\big], \\
    \boldsymbol{\Delta_k}(\mathbf{g}; \mathbf{x}_k) &= \sum_{h=0}^{w} \big[g_h\big(\mathbf{x}^{(h)}(t_k)\big) - g_h\big(\mathbf{x}^{(h)}(t_k^*)\big)\big]
\end{align*}
are the differences between the functions evaluated at the event and at the corresponding non-event, respectively, and $x^{(0)}(t) = t$. The expression (\ref{eq:bin_gam_PL}) can indeed be recognized as the likelihood of $n$ independent Bernoulli variables $Y_k=1$ with probability of success $\pi_k$ given by
\begin{align} 
\label{eq:logit_model}
\text{logit}(\pi_k) =  \boldsymbol{\Delta_k}(\mathbf{f}; \mathbf{x}_{s_kr_k}) + \boldsymbol{\Delta_k}(\mathbf{g}; \mathbf{x}).
\end{align}
When the model includes linear functionals of arbitrary smooth functions, $\boldsymbol{\Delta_k}(\mathbf{f}; \mathbf{x}_{s_kr_k})$ and $\boldsymbol{\Delta_k}(\mathbf{g}; \mathbf{x}_k)$, the expression (\ref{eq:bin_gam_PL}) is the likelihood of a degenerate logistic additive model. This means that efficient implementations from the generalised additive modelling literature \citep{hastiegeneralized,woodgeneralized} can now be used to fit flexible smooth effects for all covariates, including the global ones. 

Two points are worth mentioning. Firstly, the degeneracy resulting from the likelihood only being composed of successes does not pose a problem for the estimation, since no intercept is included in the model. Secondly, the linear functionals involve the difference of functions at different covariate values from two different time points. Within a generalized additive modelling fitting, this effectively
results in the difference of the basis functions evaluated at these two different points as being the covariates with respect to which the resulting over-parametrized generalized linear model is fitted subject to a penalization that enforces smoothness \citep[Section 6.1]{woodgeneralized}. 

\begin{figure}[tb!]
    \centering
    \includegraphics[scale=0.3]{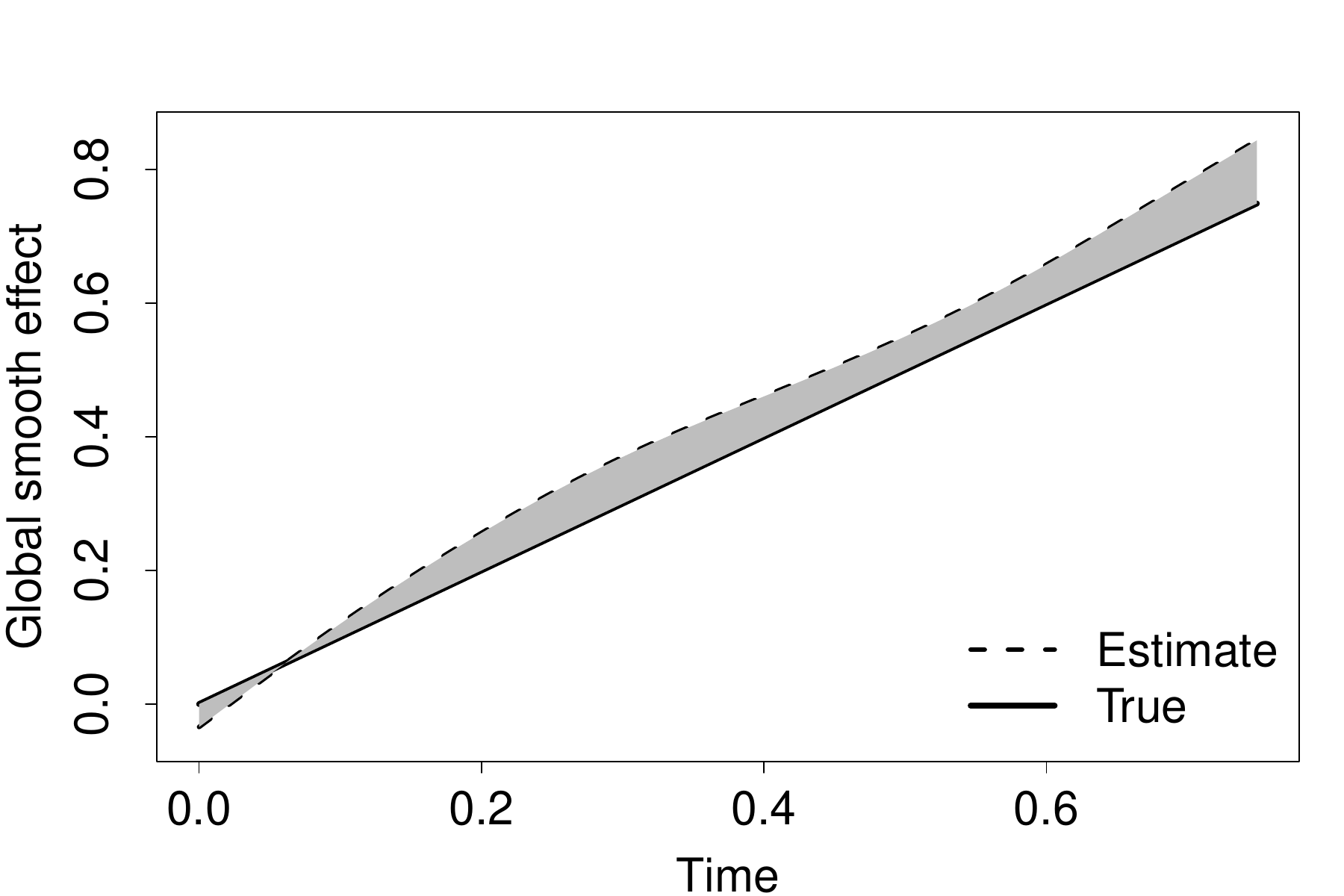}
    \caption{Example of an estimate obtained using the proposed approach (dashed curve) of a global time effect $g_0(t) = t$ (solid line). Grey area represents the divergence between the true and estimated effects, measured by the L2-norm of their difference.}
    \label{fig:single_sim_example}
\end{figure}

Maximization of the partial likelihood (\ref{eq:final_PL}) leads to estimated smooth effects for all covariates in the model, namely estimates of $\{f_l\}_l$, $\{g_h\}_h$ and $g_0$ in (\ref{eq:lambda_sr_gc}). 
Figure \ref{fig:single_sim_example} shows an example of one such estimate obtained for data in which the global time effect is
$g_0(t) = t$. 
The formulation of the model (\ref{eq:logit_model}) means that the effects are identifiable only up to an additive constant, and thus that the intensity process (\ref{eq:lambda_sr_gc}) can only be recovered up to the multiplicative constant $\lambda_0$. In the generalized additive modelling implementations, this is handled using an appropriate re-parametrization by means of a centering constraint \citep[Section 5.4.1]{woodgeneralized}, leading to estimates of smooth terms which are centered around zero. 
If one is interested in recovering the underlying speed of the event process, the constant $\lambda_0$ can be estimated by fitting a linear model to the values of the Breslow estimator at the event times. Indeed, after having accounted for all the estimated effects,  the residual cumulative baseline hazard is a linear function of time passing through zero with slope equal to $\lambda_0$.  
In the special case when there is only one global time effect, one can use this estimate of $\lambda_0$ to appropriately rescale the plot of the smooth effect and effectively compare it to the true one. 

\subsection{Shift distribution and estimation uncertainty}
\label{subsec:var_shift}
The shift distribution in principle does not affect the estimation of the dyadic covariates $\mathbf f$, but it does affect the precision of the estimates the global covariates. Clearly, a zero shift would result in an infinite variance. In this section we investigate the relation between the shift distribution used, the variance of the global covariates and the resulting uncertainty around the estimated effects $\mathbf g$. 

Each effect included as a smooth term in (\ref{eq:lambda_sr_gc}) is expressed as a linear combination of a particular set of basis functions $\{b_j\}_{j=1}^B$. For simplicity, we consider a single global covariate $x(t)$ and no dyadic effects,
$$\lambda_{sr}(t) = \lambda_0  \exp\left\{\sum_{j=1}^B \theta_j b_j(x(t))\right\}.$$ 
In order to evaluate the variance of $\theta_j$, we consider one single observation of an event $(t,s,r)$ and its corresponding sampled non-event $(t^*,s^*,r^*)$. The observed information, given as the second derivative of the log-likelihood $\log{\mathcal{L}^{PS}}$ with respect to $\theta_j$, is given by
\begin{equation*}
    -\frac{\partial^2 \log{\mathcal{L}^{PS}}}{\partial \theta_j^2}  = 
  \big(b_j^2(x(t))\pi + b_j^2(x(t^*))(1-\pi)\big) - \big(b_j(x(t))\pi + b_j(x(t^*))(1-\pi)\big)^2= \mathbb{V}  B_j,
\end{equation*}
where $\pi = \exp\{\sum_j \theta_j b_j(x(t))\}/(\exp\{\sum_j \theta_j b_j(x(t))\} + \exp\{\sum_j \theta_j b_j(x(t^*))\})$. The information about $\theta_j$ can be seen as the variance of a random variable $B_j$ taking value $b_j(x(t))$ with probability $\pi$ and $b_j(x(t^*))$ with probability $1-\pi$. This is directly related to the variability of the shifts, when we consider the definition of the non-event time with respect to the event time, namely $t^* = t + h_{sr} - h_{s^*r^*}$. Given the continuity of the basis functions, the information increases when the shift $\Delta h = \abs{h_{sr} - h_{s^*r^*}}$ increases. From this insight, we conclude that the shift distribution should be large. We will validate this empirically in the simulation study, as well as checking the robustness of the estimation when the average shift size increases.

\section{Simulation study}
\label{sec:simulation}

In this section, we investigate the effectiveness of the proposed approach by means of several simulation studies. In Section \ref{subsec:n_p_nu_sim}, we assess the role that sample size, network size and choice of shift distribution have on the performance of the proposed method. Then, in Section \ref{subsec:poisson_FL_sim}, we provide a comparison between the proposed method and the main competitor, the full likelihood approach proposed in \cite{stadtfeld_Block_global_FL}.

\subsection{Effects of sample size, network size and shift distribution}
\label{subsec:n_p_nu_sim}
In this section we evaluate the efficacy of our method by varying the sample size (number of events), network size (number of nodes) and shift distribution (average shift) in a simulation study. We consider a variety of covariates, global and non-global, node-specific and dyadic, exogenous and endogenous, time dependent and not. In particular, letting $\mathcal{S} = \mathcal{C} = \{1,\dots,p\}$, we consider the following intensity process
\begin{align}
    \label{eq:lambda_sr_sim}
    \lambda_{sr}(t) = 
    \begin{cases}
        \exp\{t + \beta_1 x_s + \beta_{2} x_{sr} + \beta_{\text{rep}}x_{\text{rep},sr}(t) +\beta_{0}x(t) \} ,& \; s\neq r\\
        0,& \; s=r
    \end{cases}
\end{align}
where $g_0(t) = t$ is the global time effect, $x_{\text{rep},sr}(t)$ is a binary endogenous \emph{repetition} variable representing whether the pair $(s,r)$ occurred at least once up to time $t$, $x_s$ is an exogenous variable associated to the sender, with $x_s \sim N(5,1)$, for $s = 1 \ldots, p$, $x_{sr}$ is a dyadic exogenous covariate defined by $x_{sr} = |x_s-x_r|$, while $x(t)$ is a global covariate defined as a time-dependent and periodic, piecewise constant function (exact definition can be found in the Supplementary Materials). The true regression coefficients are set to
\[
\beta_1 = 0.5, \ \beta_{2} = -1, \ \beta_{\text{rep}} = 1.5, \ \beta_{0} = -0.7.    
\]
To sample relational events according to the intensity processes $\lambda_{sr}$ defined above, we simulate the counting process $\mathbf{N}$ as an inhomogeneous Poisson process via the \textit{$\tau$-leap algorithm} \citep{tau-leap}. A detailed description of how we use this algorithm can be found in the Supplementary Materials. Briefly, inter-arrival times between successive events are generated 
from an exponential distribution by assuming a constant rate in the corresponding infinitesimal time interval of length $\tau$. 
For each event time obtained in this way, we then use a multinomial distribution to sample  the pair that is due to occur from the risk set. In the sampling, we do not allow self-loops and assume that all other pairs are constantly at risk of happening.  

For the construction of the time-shifted event process, we consider exponentially distributed shifts with a mean proportional to the average simulated event time according to a proportionality constant $\nu$.
Estimation of the parameters via the partial likelihood (\ref{eq:final_PL}) is performed using the \texttt{gam} function from the R package 
\texttt{mgcv}, using regression terms for each covariate and a thin plate regression spline of rank 10  for the global time effect $g_0$ \citep{woodtprs}. Mean-prediction error is used for the optimal choice of the smoothing parameter. 

We fix the sample size to $n=3000$ events, the network size to $p=15$ nodes and the mean shift equal to the average simulated event ($\nu=1$). We then consider 3 settings where we vary one of these parameters in turn, while keeping the other two fixed. For each setting, we perform 100 simulations from the process and summarise the results in terms of estimation of the regression coefficients of the individual covariates $\beta$ and of the residual global time effect $g_0(t)$. For the latter, we calculate the $L2$-norm of the difference between the true and estimated effect over the observed time period.


\paragraph{Effect of number of events.}
\begin{figure}[tb!]
    \centering
    \begin{subfigure}{.49\linewidth}
    \centering
    \includegraphics[scale=0.25]{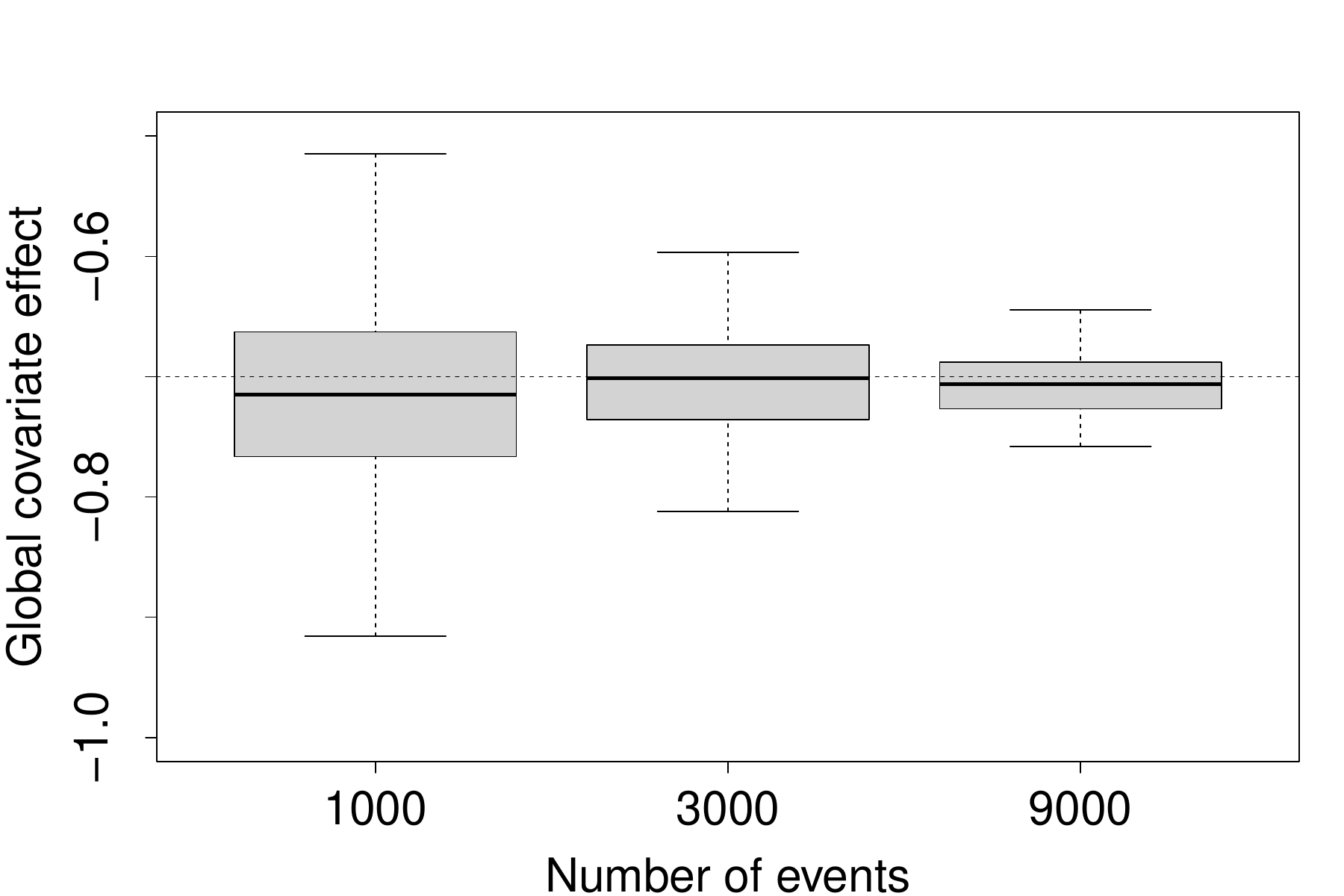}
     \caption{}
    \label{fig:different_n_gc}
    \end{subfigure}
    \hfill
    \begin{subfigure}{0.49\linewidth}
		\centering
    \includegraphics[scale=0.25]{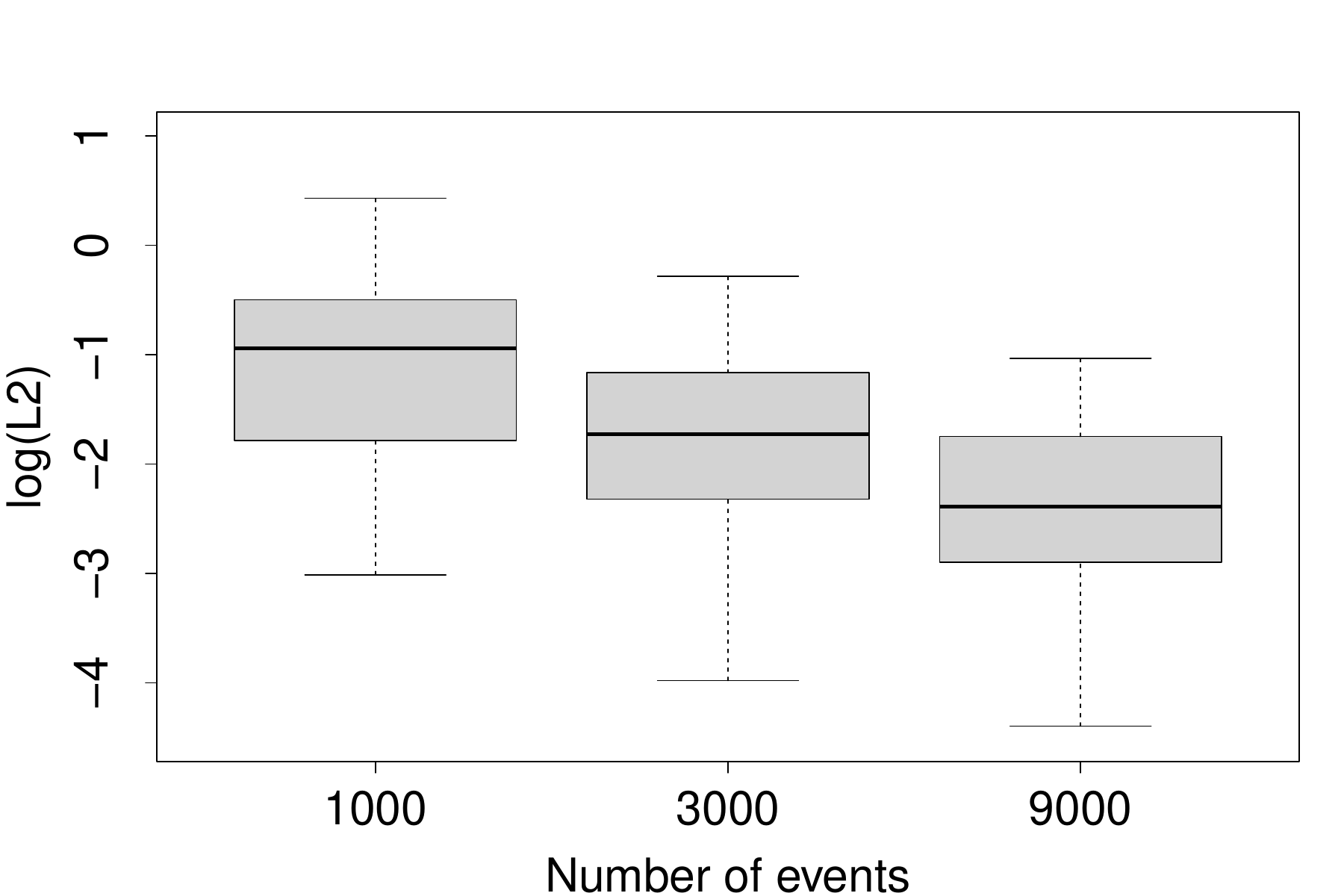}
     \caption{}
    \label{fig:different_n_l2_gte}
    \end{subfigure}
    \caption{Results across 100 simulations with $p=15$, $\nu=1$ and $n \in \{1000, 3000, 9000\}$. As the number of events $n$ increases, (a) the global covariate effect $\beta_{0}$ is centered around the true value (dotted line) and its precision increases  (b) the L2-norm (in log-scale) of the difference between the true and estimated global time effect $g_0(t)$ decreases. }
\label{fig:different_n}
\end{figure}
Figure \ref{fig:different_n} shows the estimation results for increasing values of the number of events $n \in \{1000, 3000, 9000\}$ across 100 replications. Figure \ref{fig:different_n_gc} shows a boxplot of the estimates of the regression coefficient of the global covariate $\beta_{0}$.  The figure shows how the estimates are centered around the true value (dotted line) and the uncertainty around the estimates decreases as the number of events increases. Similar plots are obtained for the other covariates and are reported in the Supplementary Materials. A similar behavior is observed in Figure \ref{fig:different_n_l2_gte} for the residual global time effect $g_0(t)$. As $n$ increases, the estimated function approaches the true function, as shown by a decreasing L2-norm of the differences between the two functions. 

\paragraph{Effect of number of nodes.}
\begin{figure}[tb!]
    \centering
    \begin{subfigure}{.49\linewidth}
        \centering
         \includegraphics[scale=0.25]{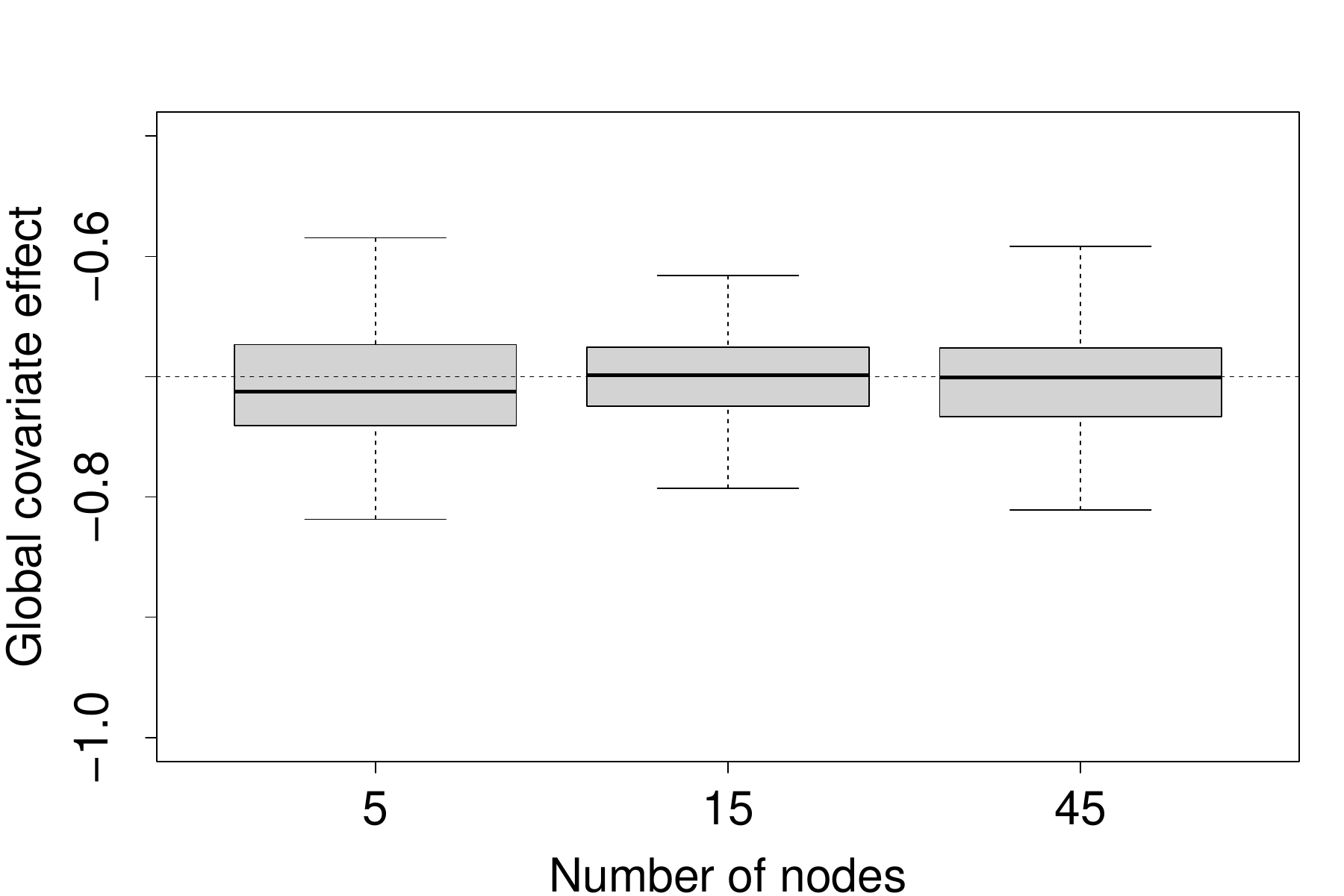}
         \caption{}
         \label{fig:different_p_gc}
    \end{subfigure}
    \hfill
    \begin{subfigure}{0.49\linewidth}
        \centering 
				\includegraphics[scale=0.25]{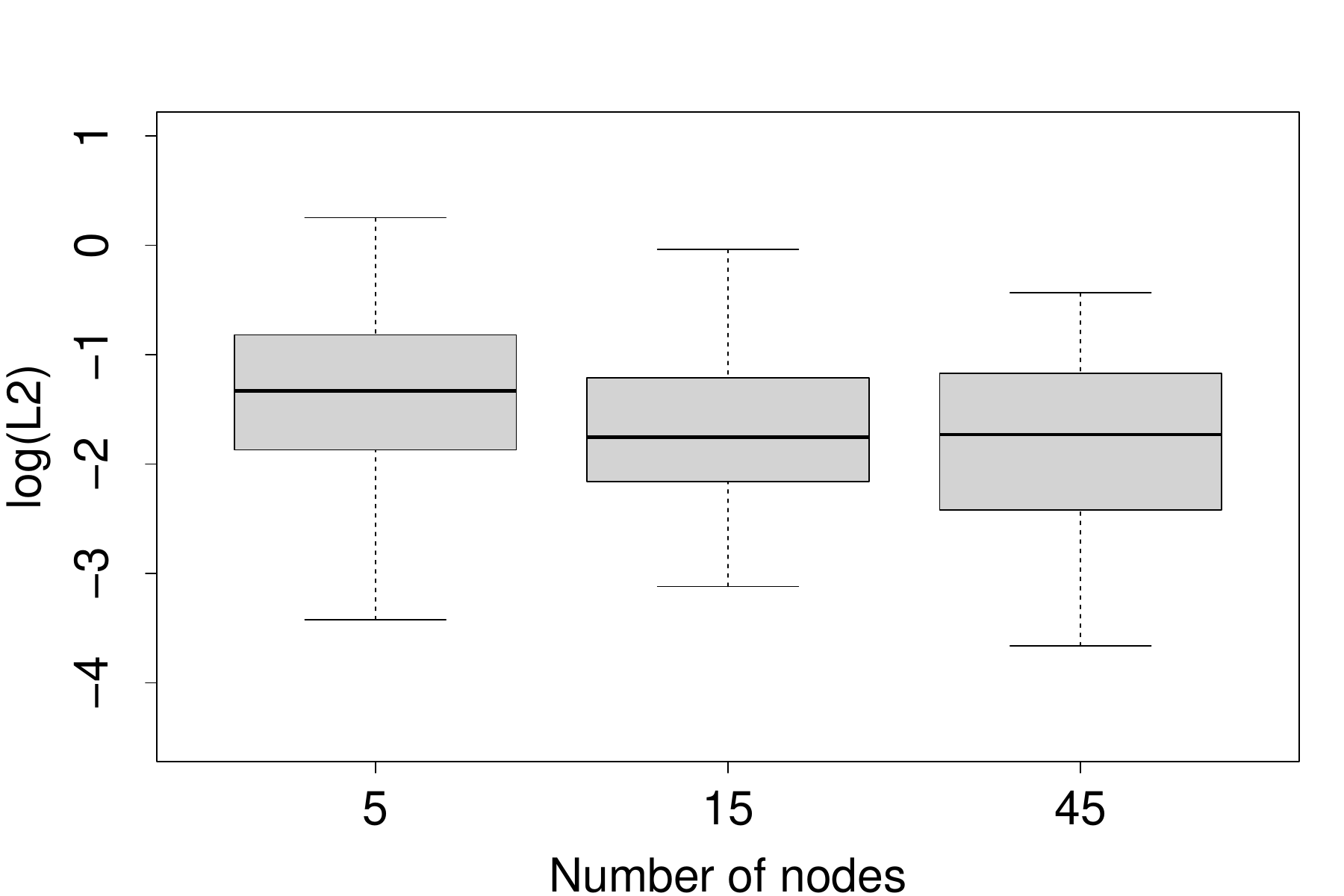}
         \caption{}
         \label{fig:different_p_l2_gte}
    \end{subfigure}
    \caption{Results across 100 simulations with $n=3000$, $\nu=1$ and $p \in \{5, 15, 45\}$. An increasing number of nodes $p$ does not have a strong effect on either (a) estimation of the global covariate effect $\beta_{0}$ or (b) estimation of the global time effect $g_0(t)$.}
\label{fig:different_p}
\end{figure}
Figure \ref{fig:different_p} shows the results for increasing values of the number of nodes $p \in \{5, 15, 45\}$ across 100 replications. Differently to the previous results, we find no strong effect of the number of nodes on either the estimates of the global covariate effect in Figure~\ref{fig:different_p_gc} or of the global time effect in Figure~\ref{fig:different_p_l2_gte}. This is to be expected as the number of nodes neither increases the information in the data, nor the model complexity of the generating system.

\begin{figure}[tb!]    
	\centering
	\begin{subfigure}{.49\linewidth}
		\centering
		\includegraphics[scale=0.25]{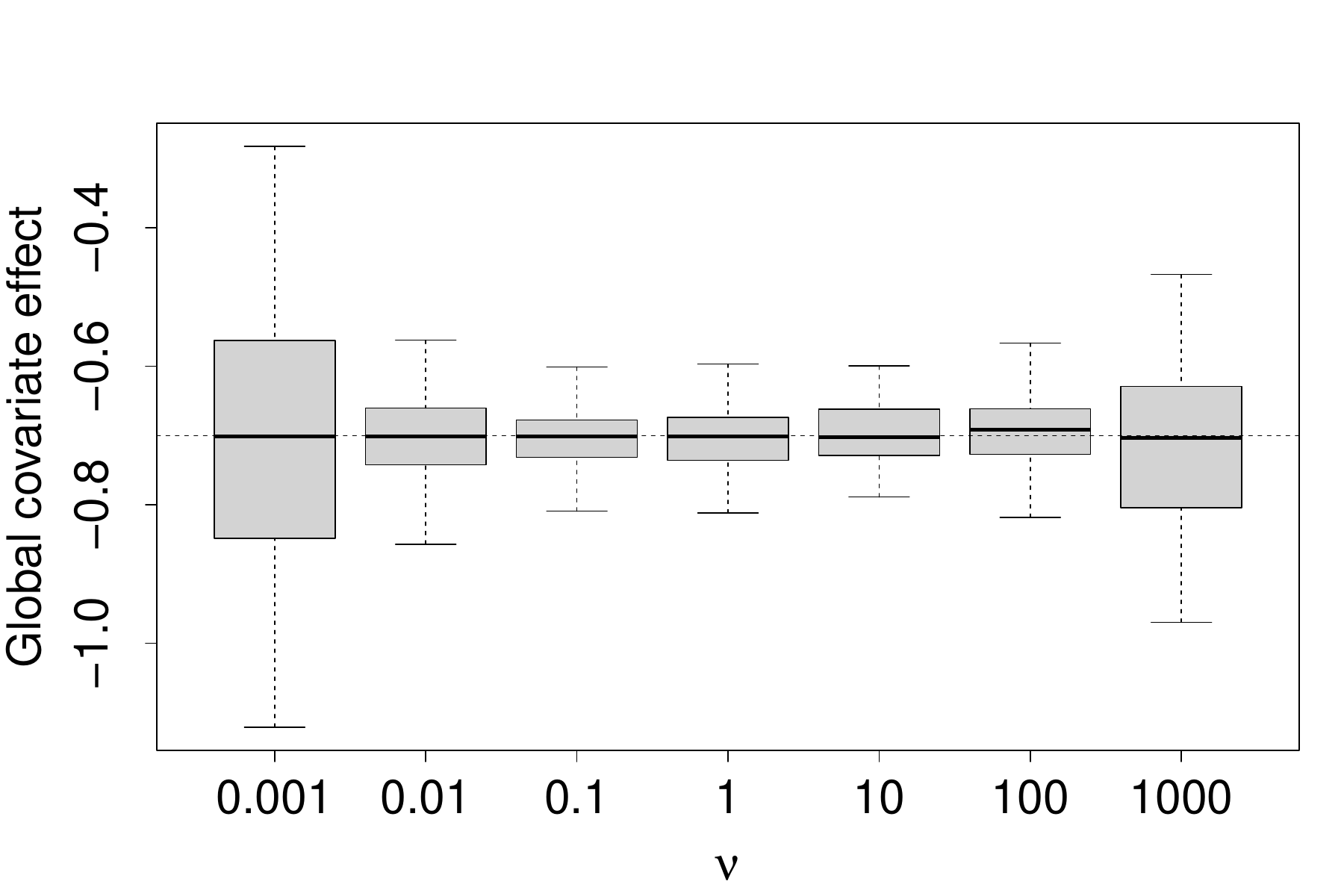}
		\caption{}
		\label{fig:different_mf_gc}
	\end{subfigure}
	\hfill
	\begin{subfigure}{0.49\linewidth}
		\centering 
		\includegraphics[scale=0.25]{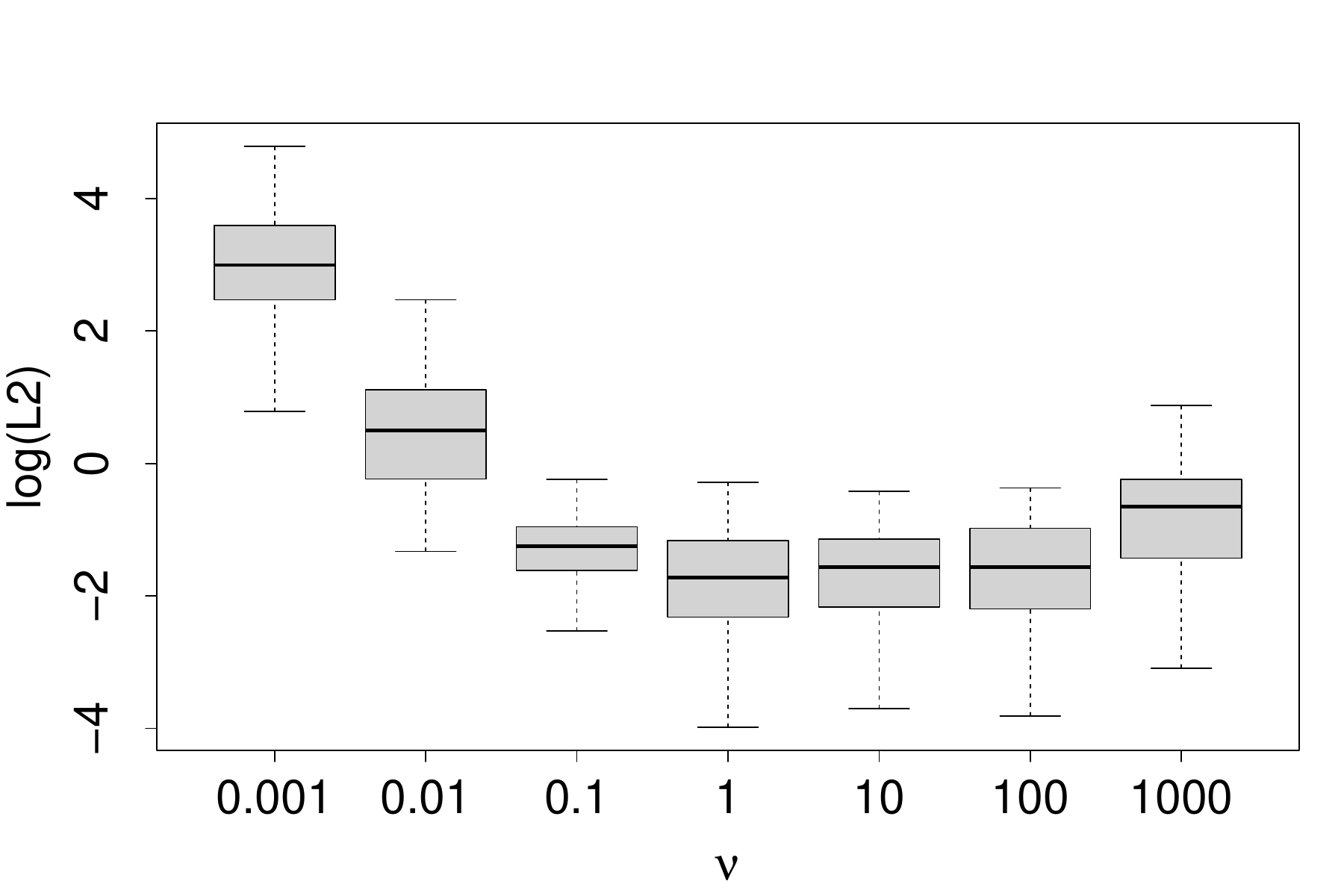}
		\caption{}
		\label{fig:different_mf_l2_gte}
	\end{subfigure}
	\caption{Results across 100 simulations with $n=3000$, $p=15$ and $\nu \in [0.001,1000]$. Too small or too large shifts have a negative effect both on (a) estimation of the global covariate effect $\beta_{0}$ and (b) estimation of the global time effect $g_0(t)$. No effect for intermediate values of $\nu \in [0.1,100]$.}
\label{fig:different_mf}
\end{figure}

\paragraph{Effect of shift distribution}
Figure~\ref{fig:different_mf} shows the results for increasing the shift distributions by changing the values of $\nu \in [0.001, 1000]$. Figure~\ref{fig:different_mf_gc} shows good results for intermediate values of $\nu$ but a higher uncertainty in the estimation of $\beta_{0}$ for the smallest and the largest average shift value. A similar behaviour is observed for the estimation of the global time effect (Figure~\ref{fig:different_mf_l2_gte}), with  an estimation error that decreases the greater the shifts are on average, but that starts to deteriorate for too large shifts. The reason behind this behaviour at the two extremes is different in the two cases. For too small shifts, the behaviour is supported by the theoretical results investigated in Section~\ref{subsec:var_shift} and is particularly accentuated by the temporal resolution used to define the piecewise constant global covariate $x(t)$. Indeed, for too small shifts, the event and non-event times are frequently so close to each other that $x(t)$ has the same values when evaluated at these two time points. This results in many structural zeros in the likelihood when considering the difference between the covariate of the event and of the non-event, which negatively impacts the ability of the method to effectively estimate $\beta_{0}$.
On the other hand, when shifts are too large, it occurs very often (roughly for 80\% of the observations in the simulation study) that the risk set in the shifted process at a certain event time is only composed of the pair occurring, making it impossible to sample a non-event and making a large percentage of the observations  uninformative when it comes to estimation. Such a situation clearly results in a considerable reduction in the effective sample size, which negatively impacts the precision of the estimation procedure.

\subsection{Comparison with a full likelihood approach}
\label{subsec:poisson_FL_sim}
Estimation of both global and non-global effects can also be achieved by maximizing the full log-likelihood,
\[ \ell(\beta) = \sum_{i=1}^n \log \lambda_{s_ir_i}(t_i) - \sum_{sr \in \mathcal{R}(t_i)}\int_{t_{i-1}}^{t_i} \lambda_{sr}(t)~dt.\]
For a generic observation happening at time $t_i$, this involves a sum of the integral
between $t_{i-1}$ and $t_i$ of the intensity for each pair in the risk set . This becomes quickly intractable unless some assumptions are made that simplify the integration of the hazard term. An effective solution is presented in \cite{stadtfeld_Block_global_FL} by approximating the original intensity processes as a piecewise constant function between consecutive event times, essentially assuming a piecewise exponential model. In this section, we compare the performance of this method to our method.

We simulate relational data between $p = 5$ nodes such that all possible pairs between them (including self loops) are always at risk according to the hazard of a Weibull distribution with scale parameter 1 and shape parameter $\kappa=0.1$, namely
\[    \lambda_{sr}(t) = 
        \exp\{\log(\kappa) + (\kappa-1)\log(t)\} \]
for $t \in \mathbb{R}_+$. We simulate $n\in [100, 5000]$ events. Further details on how the data are simulated can be found in the Supplementary Materials. Our partial likelihood method estimates the likelihood up to a proportionality constant, here $\log(\kappa)$. We use the global covariate $\log(t)$ and estimate the slope parameter $\kappa-1=-0.9$ both using our method and the approximate full likelihood approach by \cite{stadtfeld_Block_global_FL}. We further apply the bias reduction technique of \cite{Cordeiro_McCullagh_1991_bias_red_glm} to the estimates from both methods, using the implementation in the \texttt{brglm2} package in \texttt{R} \citep{brglm2_package} The simulations are repeated 5,000 times to obtain confidence bounds on the estimated parameters.

\begin{figure}[tb]
    \centering				
         \includegraphics[scale=0.3]{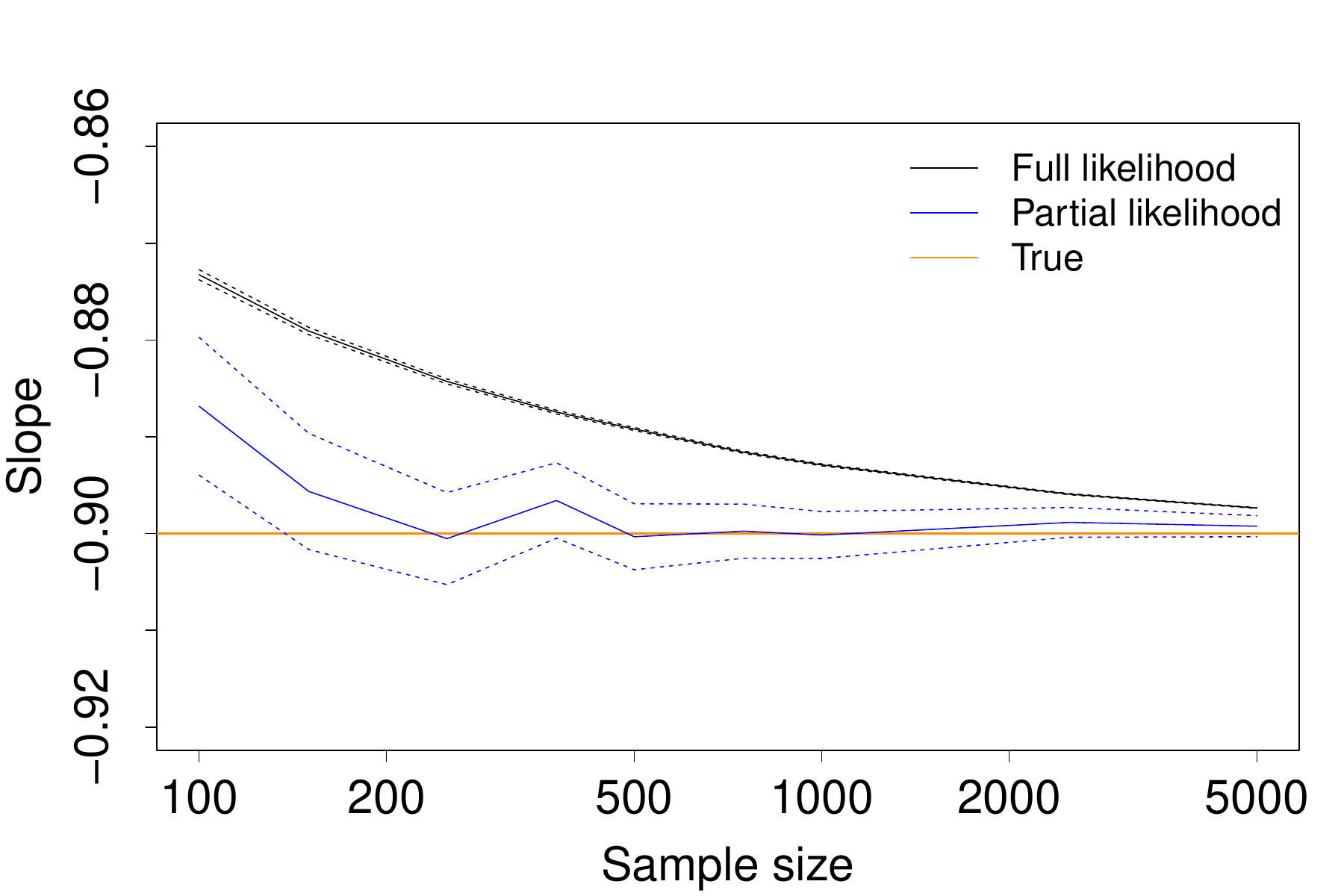}
    \caption{Average bias-corrected estimate (solid lines) and 95\% confidence bounds (dashed lines) for the slope parameter $\kappa - 1$ as sample size $n$ increases over 5000 replications. The full-likelihood estimate (black) is more precise but has higher bias than our proposed approach (blue); the latter has greater variability but is closer to the true value (orange).}
\label{fig:FL_comp}
\end{figure}
Figure \ref{fig:FL_comp} shows that although the approximate full likelihood approach has a lower variance than our proposed approach, it has a significantly larger bias, across all sample sizes $n$. This shows that the piecewise constant assumption may lead to non-consistent estimates, with a bias that is non-negligible even for large $n$. In addition, the full likelihood approach is more than a million times slower than our method, as a result of having to evaluate all non-event interactions at each event time. Although our approach has a lower precision, this can be arbitrarily improved by increasing the sampling of the number of non-events, which is currently set to a single case-control sample. In particular, sampling 100 non-events will yield a consistent estimate, which will be indistinguishable to the approximate full likelihood one in terms of variance and computationally 10,000 times faster.

\section{Bike sharing in Washington D.C.}
\label{sec:bike_data}
In this section, we use the proposed approach to model the bike sharing data from Washington D.C., described in Section \ref{sec:bike_data_descr}. The aim is to investigate the effect that global factors, like temperature and precipitation, in addition to a number of node-specific and dyadic covariates of interest, have on the rate of bike shares during the time period considered.

We propose the following model 
\begin{align}
    \lambda_{sr}(t) &= \lambda_0\exp\{g_0(t)+g_{\text{temp}}(x^{\text{(temp)}}(t)) + g_{\text{prec}}(x^{\text{(prec)}}(t)) + g_{\text{tod}}(x^{\text{(ToD)}}(t)) \nonumber\\
    &+ x_{s}^{\text{(comp)}}\beta + x_{r}^{\text{(comp)}}\gamma \\ 
    &+ f_{\text{dist}}(x_{sr}^{\text{(dist)}}) + f_{\text{rep}}(x_{sr}^{\text{(rep)}}(t)) + f_{\text{rec}}(x_{sr}^{\text{(rec)}}(t))\}. \nonumber
\end{align}
which includes a variety of variables --- both global and node/edge-specific --- that we describe below:
\begin{itemize}
\item \textbf{Global covariates.}
To investigate the role weather plays in the dynamics of bike shares, we include in the model a smooth effect of temperature $x^{\text{(temp)}}(t)$, measured in °C, and one for precipitation $x^{\text{(prec)}}(t)$, measured in mm and log-transformed to stabilize its variance. Since the data are available hourly, we consider these two functions as piecewise constant.
In addition to the usual global time effect $g_0(t)$, we also account for another smooth temporal effect, involving time of day. In particular, the covariate $x^{\text{(ToD)}}(t)$ returns a numeric value between 0 and 24 according to the hour of day corresponding to time $t$.
\item \textbf{Node-level covariates.}
In order to assess the impact that competition between bike stations has on the rate, we include two variables $x_{s}^{\text{(comp)}}$ and $x_{r}^{\text{(comp)}}$, which are defined as the distance between station $s$ (or $r$, respectively) and its closest bike station, measured in biking minutes.
High values for these covariates indicate that station $s$ (or similarly $r$) have less competition within the geographical area they are located in.
\item \textbf{Dyadic covariates.}
We model a smooth effect for the distance between stations $x_{sr}^{\text{(dist)}}$, measured in biking minutes and log-transformed. Additionally, we account for two network effects: a reciprocity effect, related to the tendency of observing a bike share going through a particular bike route given that one was previously observed in the opposite direction, and a repetition effect, related to the propensity of observing a bike share on a specific route, given that another one previously occurred along the same route. We do this by defining smooth effects of the following two endogenous covariates
\begin{align*}
    x_{sr}^{\text{(rep)}}(t) &= \exp\{-\delta_{sr}^{\text{(rep)}}(t)/2m_{\text{rep}}\},~~\delta_{sr}^{\text{(rep)}}(t) = \min_{t_k<t;s_k=s;r_k=r} t-t_k  \\
    x_{sr}^{\text{(rec)}}(t) &= \exp\{-\delta_{sr}^{\text{(rec)}}(t)/2m_{\text{rec}}\}, ~~\delta_{sr}^{\text{(rec)}}(t) = \min_{t_k<t;s_k=r;r_k=s} t-t_k 
\end{align*}
where $\delta_{sr}^{\text{(rep)}}$ and $\delta_{sr}^{\text{(rec)}}$ are the elapsed times since the same previous event or the reciprocal event to $(s,r)$, respectively, occurred. The values $m_{\text{rep}}$ and $m_{\text{rec}}$ are the medians of $\delta_{sr}^{\text{(rep)}}$ and $\delta_{sr}^{\text{(rec)}}$, respectively. The exponential transformation guards against the case of elapsed times being infinity, when the relevant event was never observed in the past, while the median makes sure that $\delta$ quantities that are not infinity do not end up being mapped to 0. The resulting variables have values between 0 and 1; the closer to zero, the further apart in time the previous same or reciprocal event is from the current event under consideration, while events repeated or reciprocated closer in time correspond to values for these variables close to 1.
\end{itemize}

For the construction of the time-shifted event process, we consider exponentially distributed shifts with a mean equal to the average event time ($\nu=1$).
Estimation of the parameters via the partial likelihood (\ref{eq:final_PL}) is performed using the \texttt{gam} function from the R package 
\texttt{mgcv}. All smooth effects are handled using thin plate regression splines of rank 10, with the exception of repetition and reciprocity, for which we consider rank 20, and the time-of-day, for which we use cyclic penalized cubic regression splines constructed over 10 evenly spaced knots \citep{woodgeneralized}.

\begin{figure}[tb]
    \begin{subfigure}{0.49\linewidth}
        \centering
         \includegraphics[scale=0.25]{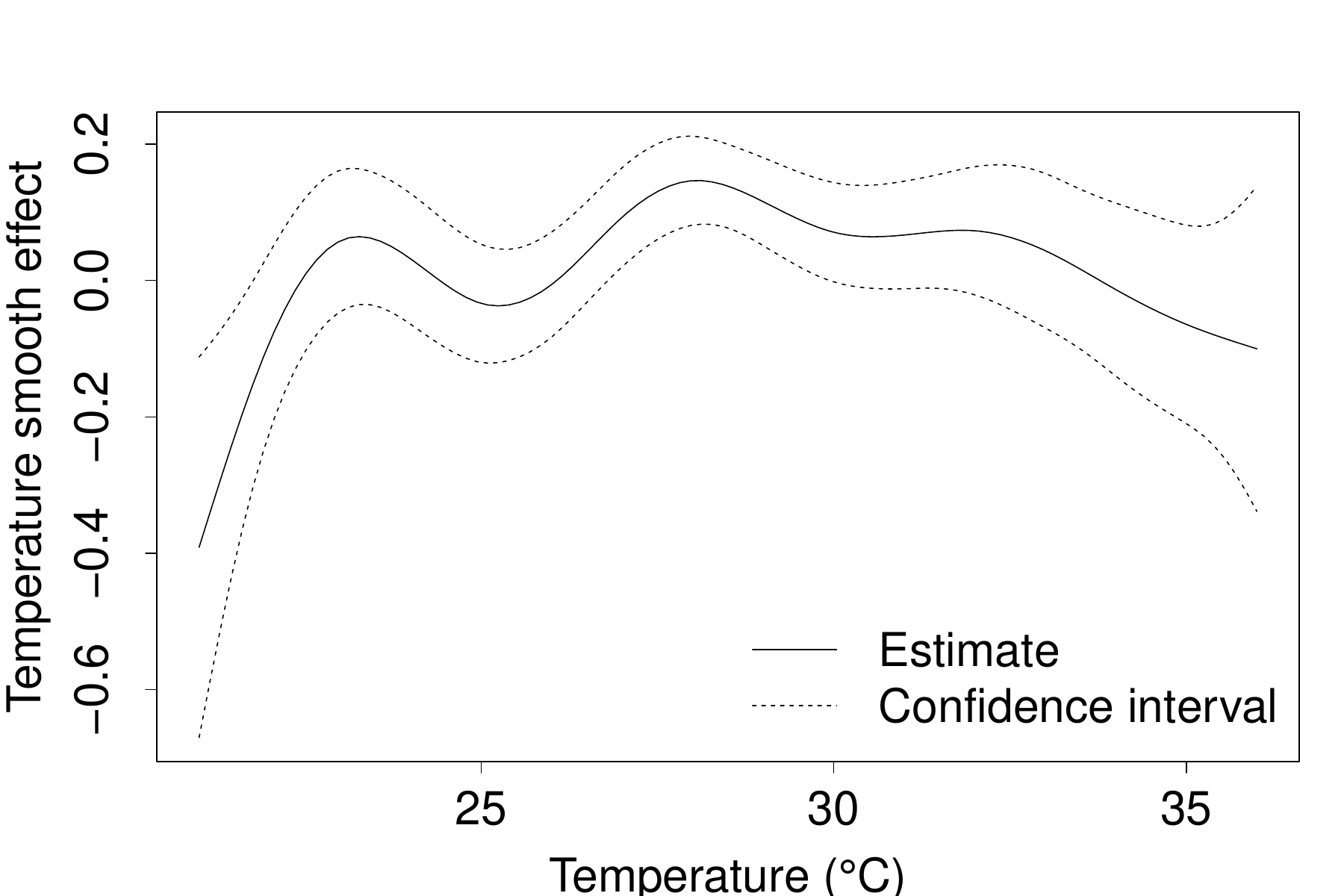}
         \caption{}
         \label{fig:bike_data_temp}
    \end{subfigure}
    \begin{subfigure}{0.49\linewidth}
        \centering
         \includegraphics[scale=0.25]{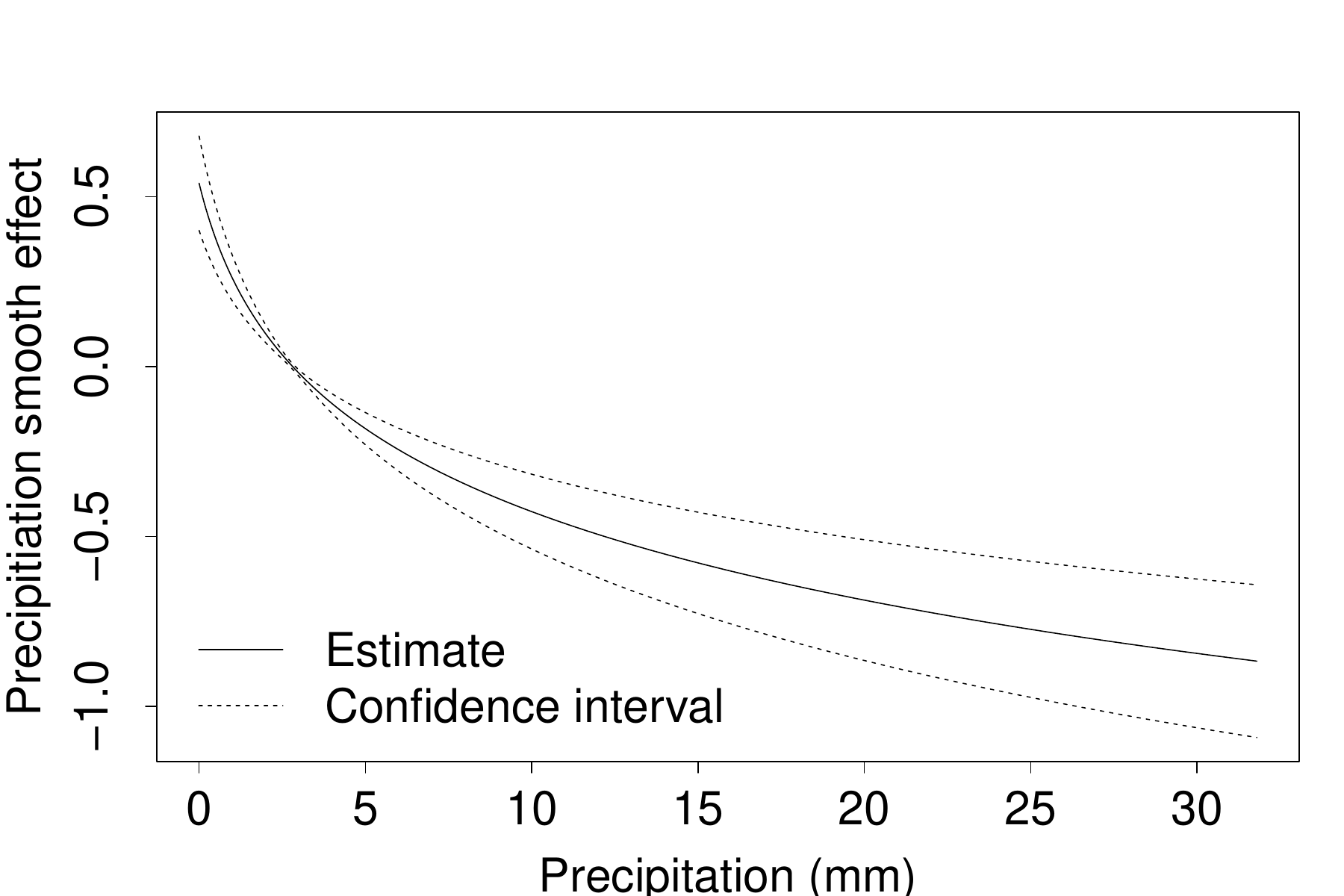}
         \caption{}
         \label{fig:bike_data_prec}
    \end{subfigure}
    \vfill
    \centering
    \begin{subfigure}{0.49\linewidth}
        \centering
        \includegraphics[scale=0.25]{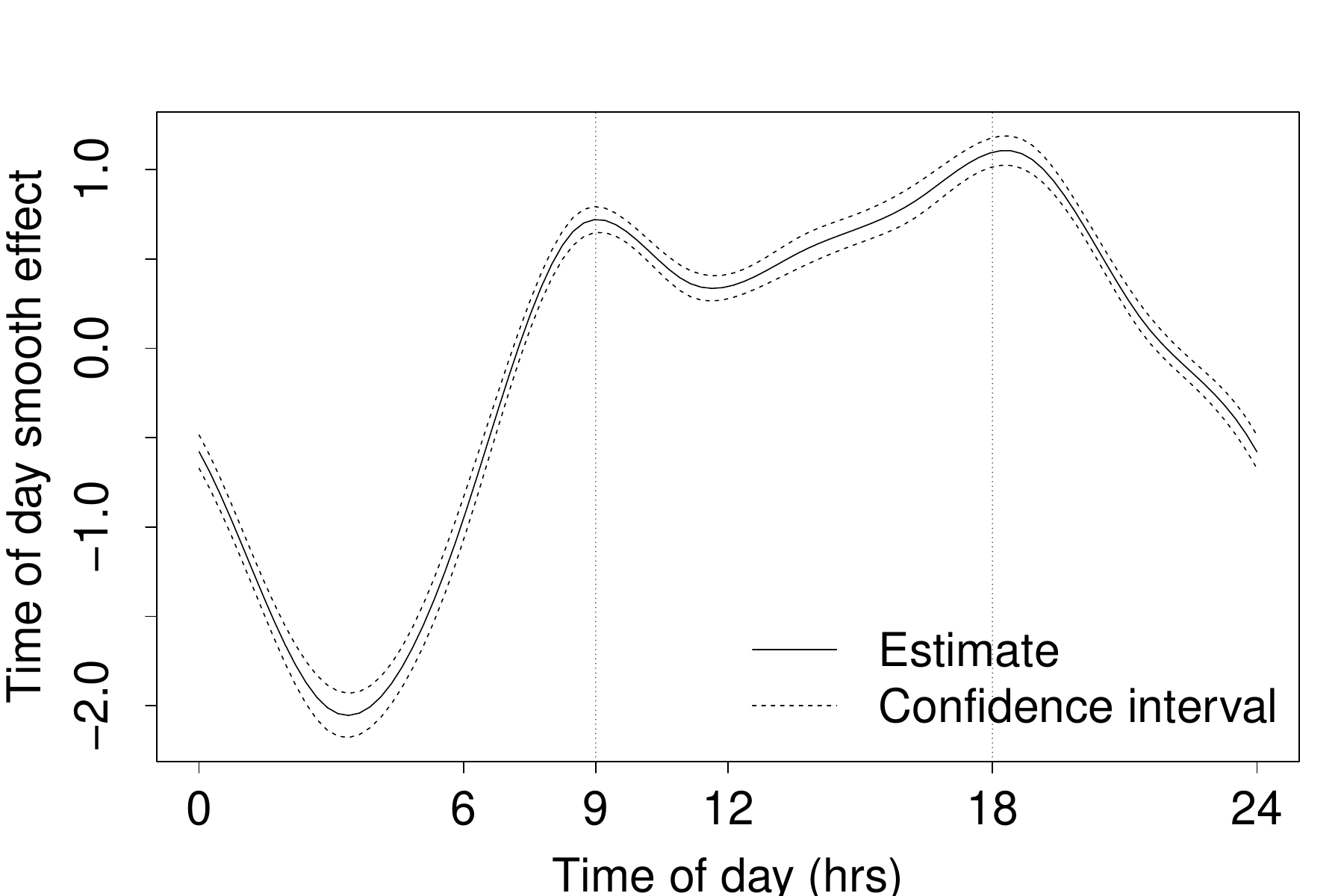}
        \caption{}
        \label{fig:bike_data_time_of_day}
   \end{subfigure} 
   \begin{subfigure}{0.49\linewidth}
   	\centering
   	\begin{tabular}{r|ccc} 
   		\hline \noalign{\smallskip}
   		& Coef. & S.E. & $p$-value  \\
   		\hline\noalign{\smallskip}
   		sender comp. $\beta$& -0.2145 & 0.0103 & $<0.0001$ \\ 
   		
   		receiver comp. $\gamma$& -0.1885 & 0.0101 & $<0.0001$  \\
   		\noalign{\smallskip}\hline
   	\end{tabular}
   	\\~~\\~~\\~~\\
        \caption{}
		\label{fig:competition}
   \end{subfigure}
   
    \caption{Effects of global and node-level covariates on bike sharing rate: (a) The warmer it is the higher the rate of bike sharing, unless it is too hot. (b) Precipitation disincentivizes bike sharing. (c) Daylight and working hours have a higher propensity to bike sharing, with an increase at the beginning (4am-9am) and at the end (6pm) of the working day. (d) Both competition parameters are negative: even though other stations are close by, this does not increase the bike share intensity, on the contrary.}
		\label{fig:bike_data_global}
\end{figure}
Figure \ref{fig:bike_data_global} plots the estimated smooth effects associated to the global covariates. As expected, Figure \ref{fig:bike_data_temp} shows how the occurrence of bike routes generally increases as the temperature increases, but starts decreasing when the temperature becomes too high. Similarly, Figure \ref{fig:bike_data_prec} shows how precipitation discourages bike sharing.
In addition to weather, time of day plays an important role in describing the tendency to rent a bike, with daylight and working hours being associated to a higher rate of bike sharing. Indeed, Figure \ref{fig:bike_data_time_of_day} shows how the intensity of bike shares between stations decreases at night (between midnight and 4am, and after 7 pm) and increases during the day, particularly between 4 and 9 am, which is most likely when people get to work, and around 6 pm, which is when they return from work.
Interestingly, the propensity to ride a bike is higher at the end of the workday time period then it is at the beginning of it.

\begin{figure}[tb]
    \begin{subfigure}{.49\linewidth}
         \centering
				        \includegraphics[scale=0.25]{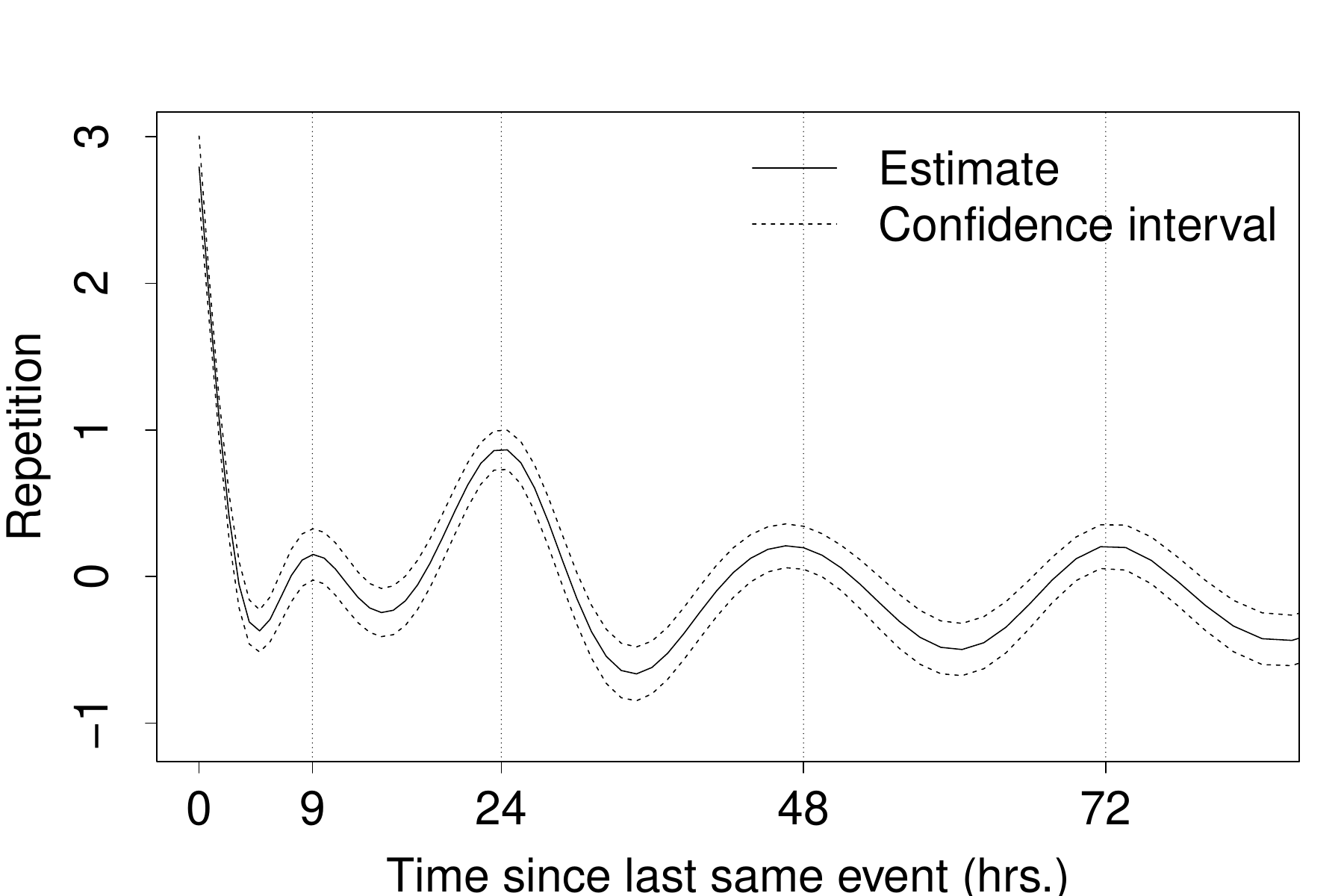}
        \caption{}
        \label{fig:bike_data_repet}
    \end{subfigure}
    \hfill
    \begin{subfigure}{0.49\linewidth}
        \centering				
         \includegraphics[scale=0.25]{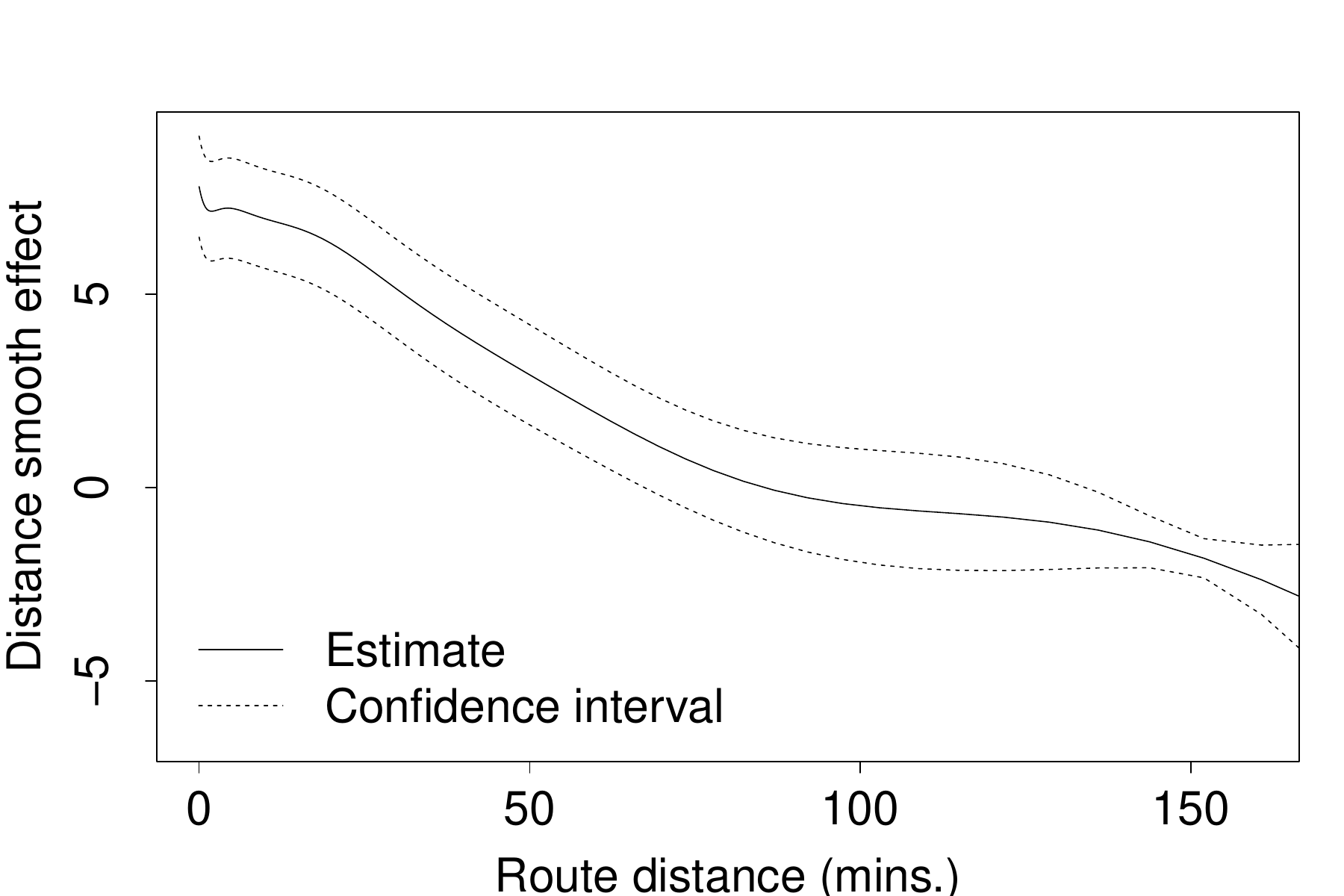}
         \caption{}
         \label{fig:bike_data_dist}
    \end{subfigure}
    \caption{Effects of 2 dyadic covariates on bike sharing rate:  (a) Bike shares have a daily repeating pattern, describing the tendency to go along the same route every day. (b) The further apart two bike stations are, the less likely is a bike share between them.}
\label{fig:bike_data_dyadic}
\end{figure}
Figure \ref{fig:bike_data_dyadic} plots the estimated smooth effects associated to two dyadic covariates, repetition and distance.   The other dyadic effect, reciprocity, as well as the global time effect, are included in the Supplementary Materials. Figure \ref{fig:bike_data_repet} shows a daily repeating pattern, with a considerable increase at 24 hours, describing the tendency to go along the same bike route on a daily basis. Interestingly, even though repetition is defined as an endogenous effect, it actually ends up capturing a daily  ``exogenous'' pattern. 
As for distance (Figure \ref{fig:bike_data_dist}), there is generally a decreasing trend as distance increases, which is to be expected, though we do not detect any decreasing effect for very short distances, where we might have expected people to not use bikes. 

In connection with the geographical location of bike stations, Figure~\ref{fig:competition} reports the regression coefficients associated to the two node-level covariates describing the sender's and receiver's competition effects, respectively. As the variables are defined as distances between a sender/receiver station and its closest station, the negative estimates suggest  a ``negative competition'' scenario. This could be explained by the lack of a sufficient number of stations for the actual volume of bike shares in the area.  Despite the close vicinity of other bike sender or receiver stations, the traffic is actually higher than for those stations that do no have other stations nearby.



\section{Conclusions} 
\label{sec:conclusions}

Relational event models provide an ideal framework for analyzing sequences of time-stamped
interactions in a variety of applied settings, including bike rides between stations. Existing inferential techniques based on the partial likelihood, only allow to investigate how node- or edge-specific covariates are able to
explain the underlying dynamics of interactions. 
Such a restriction does not allow us to completely account for the temporal nature of the data, as it is clearly possible for relevant time-dependence describing the events to not have to depend on the nodes' characteristics. 
This is particularly true when analyzing bike sharing data in which weather and temporal patterns play an important role.
In this paper, we propose an extension of relational event models with the inclusion of global covariates, that is covariates that are time-dependent but constant across interacting pairs. As the contribution of these covariates would normally drop out of the partial likelihood, we propose a time-shifted version of the original event process, from which we are able to recover the effect of all kinds of variables, including global ones. 

In order to scale the inferential approach to large dynamic networks, we propose the use of nested case-control sampling. In the specific case of one non-event sampled from the risk set at each event time, we show how the partial likelihood is that of a degenerate logistic additive model, for which efficient implementations are available that allow the inclusion of fixed, time-varying and random effects.

A simulation study shows the effectiveness
of the proposed approach in recovering both global and non-global effects and investigates the role played by sample size, network size and the shift distribution on the performance of the approach. We also provide an example that
sheds some light into the effects of the assumptions that are made when conducting inference using full-likelihood approaches and compare these to our approach. 
Finally, we show the applicability of the proposed methodology on the modelling of bike sharing data from Washington D.C., where global covariates, such as weather conditions and time of the day, are found to play an important role in describing the rate of bike sharing between two stations. 

\section{Code availability.} 
The code for reproducing the simulation study and the empirical analysis of bike sharing data can be found at the \texttt{GitHub} repository page  \href{https://github.com/MelaniaLembo/Global-covariates-REM.git}{\texttt{https://github.com/MelaniaLembo/Global-covariates-REM.git}}.

\section{Funding}
This work was supported by funding from the Swiss National Science Foundation (SNSF grant 192549).

\bibliographystyle{chicago}
\bibliography{MT_LM.bib}
\end{document}